\documentclass[fleqn,usenatbib]{mnras}

\usepackage{newtxtext,newtxmath}

\usepackage[T1]{fontenc}

\DeclareRobustCommand{\VAN}[3]{#2}
\let\VANthebibliography\thebibliography
\def\thebibliography{\DeclareRobustCommand{\VAN}[3]{##3}\VANthebibliography}

\usepackage{amsmath}
\usepackage{graphicx} 
\usepackage{subcaption}
\usepackage{float}
\usepackage{orcidlink}
\usepackage{CJKutf8} 

\newcommand{\inv}{$^{-1}$}
\defcitealias{LYS2024arXiv241119085L:I}{Paper~I} 
\defcitealias{Tremaine1984ApJ...282L...5T:TW}{TW} 

\title[Hercules: Dynamics]{On the origin of the Hercules group: II. the Trojan quasi-periodic identity on the orbital level}

\author[Li, Freeman, and Jerjen]{
Li, Yusen (李宇森)\orcidlink{0009-0003-8249-6782},$^{1}$ \thanks{E-mail: li.yusen.astr@gmail.com}
Kenneth Freeman\orcidlink{0000-0001-6280-1207},$^{1}$\thanks{E-mail: kenneth.freeman@anu.edu.au}
Helmut Jerjen\orcidlink{0000-0003-4624-9592}$^{1}$
\\
$^{1}$Research School of Astronomy and Astrophysics, Australian National University, Canberra, ACT 2611, Australia
}

\date{Accepted XXX. Received YYY; in original form ZZZ}

\pubyear{2024}

\begin{document}
\label{firstpage}
\pagerange{\pageref{firstpage}--\pageref{lastpage}}

\begin{CJK*}{UTF8}{gkai}
\maketitle
\end{CJK*}

\begin{abstract}
 The Hercules kinematic group is a stellar anomaly structure observed in the solar neighbourhood (SNd). In the previous paper, we analysed chemical signatures and related the origin of this stellar population to the outer bar. Next to consider is how this alien population migrate out into the SNd. Often, this kinematic structure is associated with bar resonances. In this paper, We consider the driving mechanism of Hercules on the orbital level. We construct a simple Milky Way-like potential model with a slowly rotating long bar and explore some of the stellar orbit families and their stability. With this model, our numerical solutions of the equations of motion show that quasi-periodic orbits trapped around fast-rotating periodic Trojan orbits around the L4 Lagrange point of the bar minor axis can pass through the SNd. When observed in the SNd, they populate the Hercules structure in the $L_Z$-$V_R$ kinematics space. Moreover, the variation in radial coverage in the galactic plane with the SNd kinematics shows good agreement with chemical signatures found in Paper I. Furthermore, the effective potential shows the topology of a volcano, the rim of which limits most orbits to stay inside or outside. Trojan orbits are a stable orbit family that can transport inner Galactic stars out to the SNd. They can explain the stellar kinematics of Hercules, and provide a straightforward basis for its chemical properties. We support that Trojan orbits associated with the slowly rotating Galactic bar explain the Hercules structure observed in the SNd.

\end{abstract}

\section{Introduction}\label{sec:intro}

\begin{figure}
    \centering
    \includegraphics[width=\linewidth]{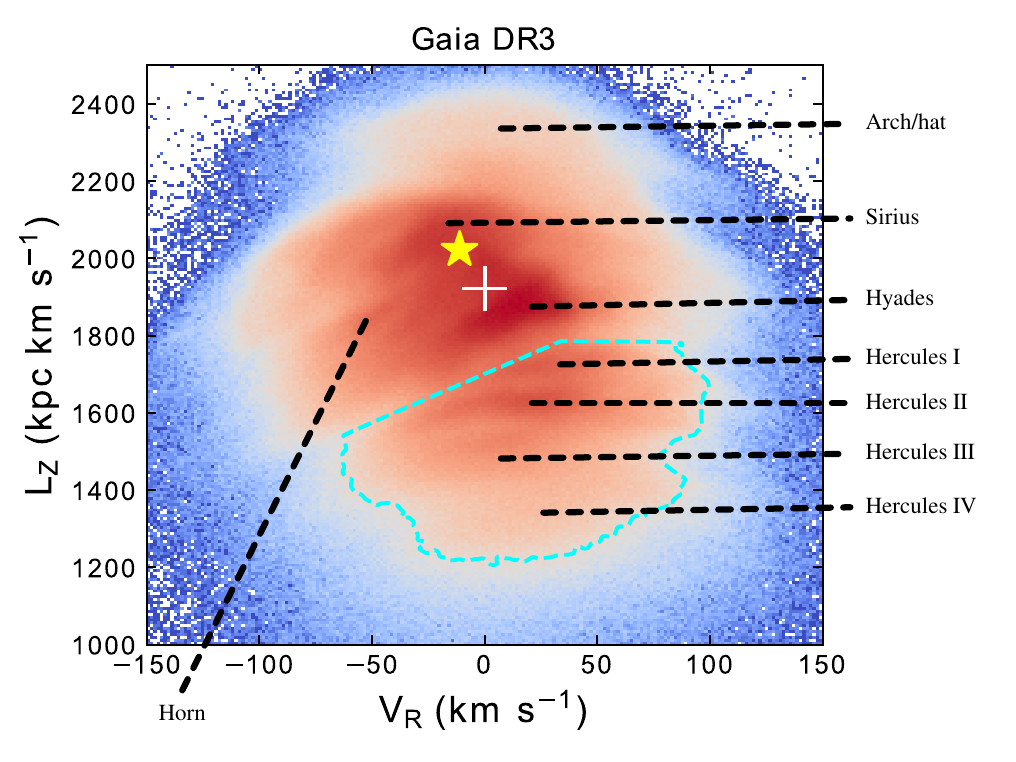}
    \caption{The distribution of more than 8 million SNd stars in \textit{Gaia} DR3 in the planar angular momentum-radial velocity plane. The yellow asterisk and the white cross mark the solar and the LSR kinematics. The Hercules group, subdivided into four subgroups, is marked by the dashed cyan contour. We associate eight over-densities as kinematic groups.}
    \label{fig:gaialzvr}
\end{figure}

The distribution of stars in the solar neighbourhood (SNd) reveals many structures in the Galactocentric cylindrical angular momentum - radial velocity ($L_Z$-$V_R$) plane as overdensities. As shown in Fig.\ \ref{fig:gaialzvr}, these structures are often referred to as \textit{kinematic groups}. Historically, due to different assumptions on the underlying mechanism, these kinematic groups have been referred to as ``\textit{moving groups}" \citep{Eggen1983AJ.....88..642E, Eggen1996AJ....112.1595E} and ``\textit{dynamical streams}" \citep{Famaey2005A&A...430..165F:stream}. 

Among the kinematic groups, one of the most extensively studied groups is the Hercules group, marked by the cyan dashed contour in Fig. \ref{fig:gaialzvr}. The Hercules group shows an asymmetry skewed toward outward radial velocity and includes angular momentum, $L_Z \sim 1300$\ to\ $1700$\ kpc\ km\ s\inv, well below the Local Standard of Rest (LSR), $L_Z \sim 1900$\ kpc\ km\ s\inv. Chemical data have identified the Hercules group as a distinctly Fe-enhanced structure in the SNd (e.g \citealp{Bensby_Feltzing2007ApJ...655L..89B, Quillen2018MNRAS.478..228Q, Khoperskov2022A&A...663A..38K}). As the metallicity of the local interstellar medium (ISM) is near-solar to subsolar \citep[e.g.][]{Pagel1981ARA&A..19...77P, Asplund2009ARA&A..47..481A}, these chemical signatures suggest an origin of the Hercules stars from outside the SNd. Our previous paper \citep[][; hereafter \citetalias{LYS2024arXiv241119085L:I}]{LYS2024arXiv241119085L:I} investigated chemical and age features with data from GALAH DR4 \citep{Buder2024arXiv240919858B} and APOGEE DR17 \citep{APOGEE2022ApJS..259...35A}. We found an iron-peak-enhanced, alpha-deficient, and odd-Z-enhanced population in the low $L_Z$ Hercules subgroups III and IV. This population is expected to originate from the outer thin bar and migrate to the SNd via some dynamical mechanism.

Due to the asymmetry in $V_R$, the Hercules group has been associated with the Outer Lindblad Resonance (OLR) of the non-axisymmetric central Galactic bar \citep{Kalnajs1991dodg.conf..323K, Dehnen2000AJ....119..800D}. This scenario is well supported by a fast rotating short bar with pattern speed $\Omega_b \sim 56$\ km\ s\inv\ kpc\inv \citep{Antoja2014A&A...563A..60A, Monari2017MNRAS.466L.113M}. However, recent photometric and spectroscopic surveys of the structure of the inner Galaxy \citep{Wegg2015MNRAS.450.4050W, Portail2017MNRAS.465.1621P} suggest a different model with a longer bar that rotates slower. The new model has a bar semi-major length $a \sim 5$\ kpc and a pattern speed $\Omega_{b} \sim 39$\ km\ s\inv\ kpc\inv. The new model places the OLR at $R \sim 10.5$\ kpc, more than $2$\ kpc outside the orbit of the sun, $R_\odot \sim 8.2$\ kpc \citep{GRAVITY2019A&A...625L..10G, GRAVITY2021A&A...647A..59G}. Hence the OLR scenario is less favoured as the Hercules origin and a revised dynamical origin is required.

Recent studies explain the origin of the Hercules group differently. \cite{Hunt2018MNRAS.477.3945H_OLR} and \cite{Asano2020MNRAS.499.2416A} explained Hercules as the 4:1 OLR of the bar. \cite{Hunt2018MNRAS.481.3794H_Transient}, \cite{Michtchenko2018ApJ...863L..37M}, and \cite{Barros2020ApJ...888...75B} explored a possible origin related to the Inner Lindblad Resonances of the spirals. \cite{Khoperskov2022A&A...663A..38K} and \cite{Liang2023ApJ...956..146L} proposed a complex scenario where the spiral arm and bar resonances both contribute to the Hercules structure and substructures. Other works attributed the Hercules structures to the corotation resonance of the long bar \citep[e.g.][]{Monari2019A&A...626A..41M, Binney2020MNRAS.495..895B, Chiba2021MNRAS.500.4710C_resonace, Chiba2021MNRAS.505.2412C_tree_ring, Wheeler2022ApJ...935...28W, Lucchini2024MNRAS.531L..14L}. Among works supporting the corotation theory, \cite{PerezVillegas2017ApJ...840L...2P} and \cite{DOnghia2020ApJ...890..117D} examined the mechanism in finer detail, from the perspective of stellar orbits. They propose orbits around the L4 and L5 Lagrange points on the bar minor axis as building blocks of the Hercules structure. These orbits are often referred to as Trojan orbits, analogous to the orbits of planetary Trojan asteroid families, which are also trapped around the L4 and L5 Lagrange points in the stellar-planar restricted three-body problem \citep{Murray1999ssd..book.....M}. However, most of the previous studies were based on N-body simulations. While simulations provide self-consistent Galaxy models, we investigate the Trojan orbit scenario at the stellar orbital level and relate the dynamical theory to footprints in chemical abundances in this paper.

We adopt a simple but realistic barred potential model and restrict our scope to motions in the galactic plane ($z = 0$) to capture some general properties of regular planar orbits in barred disc galaxies. In future studies, we plan to adopt a more realistic, observation-based, 3D potential of the Milky Way Galaxy that includes details of the inner Galaxy with a boxy peanut-shaped bulge and a super thin long bar \citep{Sormani2022MNRAS.512.1857S, Hunter2024arXiv240318000H}. We will use this more realistic model to consider possible inner galactic orbits that might be captured by the Trojan orbits and the origins of the super thin bar with a vertical scale height only about $\sim 40$ pc \citep{Wegg2015MNRAS.450.4050W}. 

In \S\ref{sec:method}, we present the analytical model of the barred galaxy potential to study the orbital dynamics. In \S\ref{sec:dynamics}, we investigate the nature of Hercules-related Trojan orbits with the gravitational model as quasi-periodic orbits around stable periodic orbit families around the stable L4 Lagrange point, and analyse the behaviours and stability of these orbits with their surfaces of section. In \S\ref{sec:discussion}, we relate orbital studies with chemical signatures found in \citetalias{LYS2024arXiv241119085L:I} and discuss possible scenarios of capturing inner Galactic stars into Trojan orbits. Finally, a summary is given in \S\ref{sec:summary}.

\section{Data and model}\label{sec:method}

\subsection{Data}\label{sec:data}

We follow the methods presented in our previous paper to obtain the kinematics data from \textit{Gaia} DR3 \citep{Gaia2023A&A...674A...1G}. We obtain the 6D phase space kinematics of proper motions, line-of-sight radial velocity, and parallax in \textit{Gaia}. By assuming a Local Standard of Rest (LSR) with a tangential velocity of $v_{T,\text{ LSR}}= 235$\ km\ s\inv, they are then transformed into Galactocentric cylindrical coordinates with angular momentum $L_Z = V_T R$ and radial velocity $V_R$, positive away from the Galactic centre. The solar neighbourhood (SNd) is defined as a cylinder with radius $R_{\text{SNd}} = 1$\ kpc and height $Z_{\text{SNd}} = 1$\ kpc centred radially and vertically at the sun.

The resulting distribution in the $L_Z$-$V_R$ plane, as presented in Fig.\ \ref{fig:gaialzvr}, shows the eight most prominent kinematic groups. We observe the high $L_Z$ Arch/Hat; the slightly negative $V_R$-biased Sirius; the most populated Hyades; the Horn on the left of Hyades; and the four Hercules subgroups I to IV. The LSR and solar kinematics are marked by the white cross and the yellow asterisk for reference. The Hercules group includes about $23$\ per\ cent of stars in the SNd under our definition. This fraction varies slightly among the surveys used (\textit{Gaia}, GALAH, APOGEE). 

\subsection{The simple dynamical model}\label{sec:model}

As the Hercules stars are defined by the pattern in the kinematics plane, we wish to understand their dynamical features. To study the stellar orbits in the Galaxy, we construct a barred analytic gravitational potential that models the Galaxy. Our barred model include our current best understanding of the Galactic bar as a long, slowly rotating bar. However, we note that different bar parameters (e.g. a faster bar) would result in models that support different results and conclusions. As an example, a scenario associated with the outer Lindblad resonance (OLR) is favoured for the formation of Hercules in a short, fast bar model \citep[e.g][]{Dehnen2000AJ....119..800D, Antoja2014A&A...563A..60A}.
 
\begin{figure*}
    \centering
    \includegraphics[width=\linewidth]{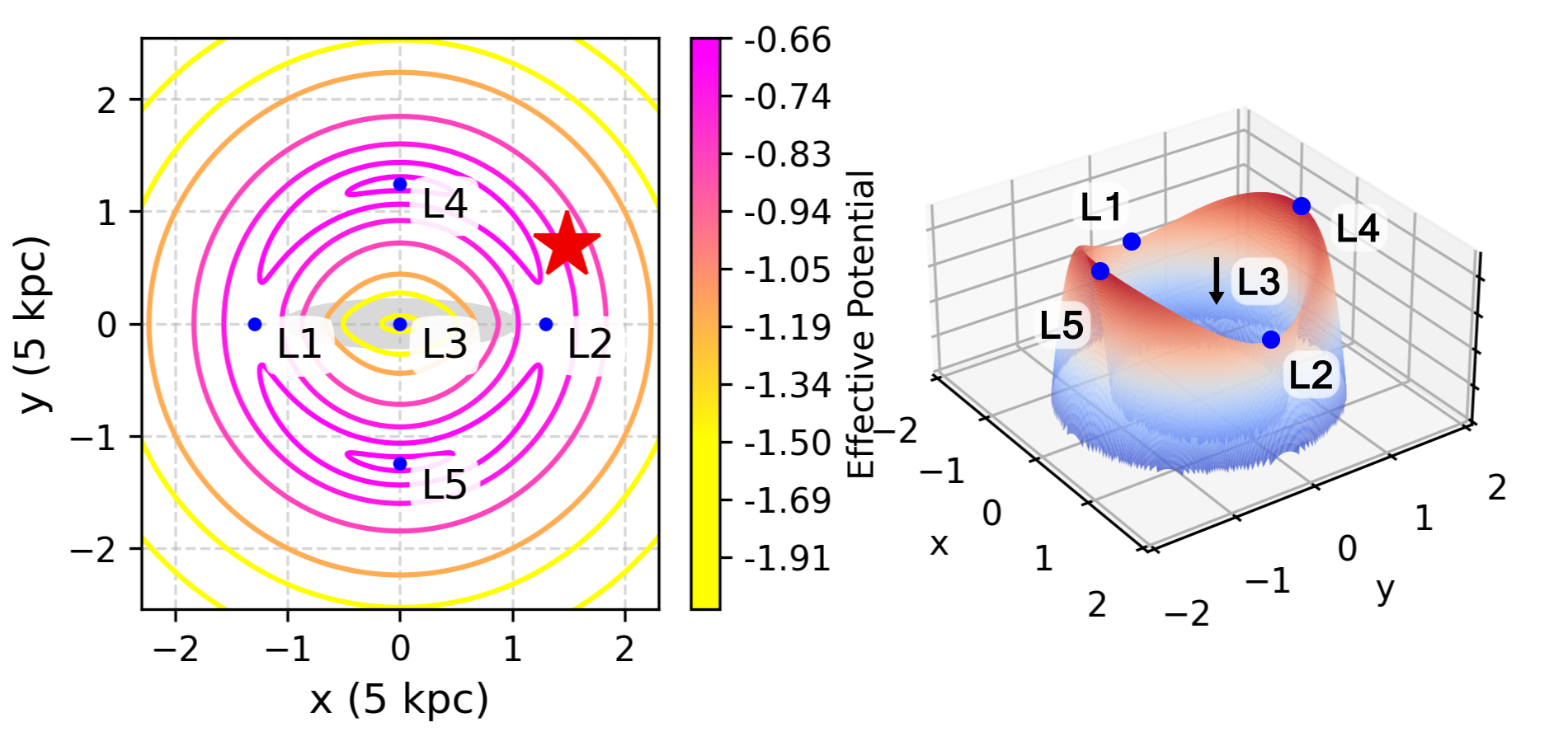}
    \caption{The effective gravitational potential of the combined model. Left: contour plot of the combined potential. The five critical points (Lagrange points) are marked by blue points. The position of the sun is marked by the red star. Right: 3D plot of the volcano-like topology of the effective potential. The rim of the volcano functions as a boundary that prevents the communication between the inner and outer region. Most orbits stay either inside or outside the rim, except Trojans and some high $E_J$ orbits.}
    \label{fig:potential}
\end{figure*}

We set our frame of reference corotating with the galactic bar, and align the x-axis with the bar major axis. We let the bar rotate clockwise as seen in Fig.\ \ref{fig:potential}, consistent with our presentation of the \textit{Gaia} kinematic data. We adopt a disc plus dark halo approximation of the Galaxy and restrict our study to the 2-dimensional Galactic plane to study regular planar stellar motions.

In this model, the conservation of angular momentum ($L_Z$) and energy ($E$) does not hold in general. Instead, the Hamiltonian in the rotating frame, or the Jacobi Integral $E_J$, is the only conserved quantity that can be written in analytic form.

\begin{equation}
    E_J=\frac{1}{2}{|\dot{\mathbf{r}}|}^2+\Phi_{\text{eff}} = E -\Omega_b L_Z,
    \label{eqn:ej}
\end{equation}

where $\Phi_{\text{eff}}=\Phi-\frac{1}{2}\Omega_b^2{|\mathbf{r}|}^2$ is the effective potential in the frame corotating with the bar rotating at an angular velocity $\Omega_b$, and $\textbf{r}\ \&\ \dot{\textbf{r}}$ are the planar position and velocity in the rotating frame. In general, analytical solutions for orbits in such rotating potentials are rare (e.g \citealp{Freeman1966MNRAS.133...47F}). Hence, we rely on numerical solutions for most of the understanding on orbital properties in this barred galaxy model.

For this purpose, an analytical non-axisymmetric bar potential $\Phi_b$ is combined with an axisymmetric component $\Phi_d$ that accounts for the distribution from the bulge, disc, and dark halo of the galaxy to simulate the flat rotation curve widely observed in many disc galaxies including the Milky Way \citep{Eilers2019ApJ...871..120E}.

\begin{figure}
    \centering
    \includegraphics[width=\linewidth]{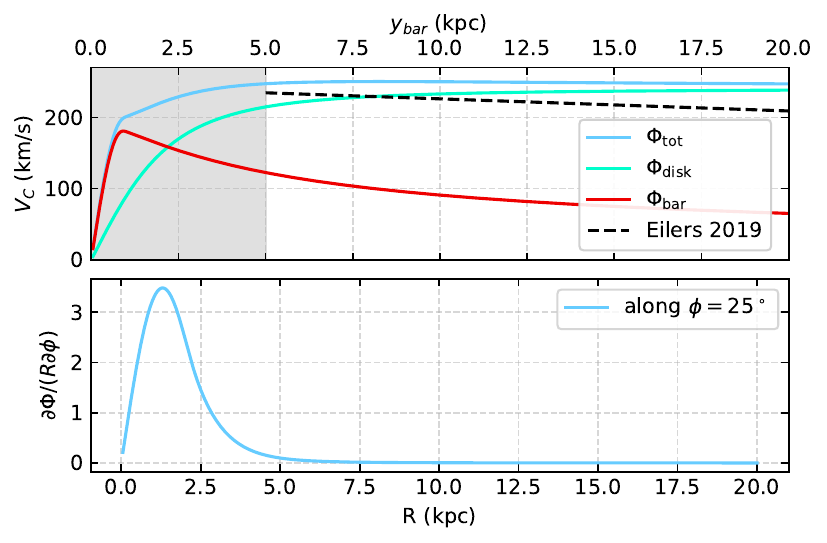}
    \caption{Rotation curve and azimuthal force induced by the potential model on stationary particles in the frame of corotation. Top: the rotation curves generated by the different components in the model along the y-axis. The observational fitted rotation curve from \protect\cite{Eilers2019ApJ...871..120E} is plotted for comparison; Bottom: the azimuthal force in the direction of the sun ($\phi = 25^\circ$) at different galactic radii in the model. The bar influences mainly the inner part of the Galaxy $R < 10$\ kpc.}
    \label{fig:rotcurve}
\end{figure}

In this rotating frame, the equations of stellar motion are,

\begin{align}
    \label{eqn:eomxd}
    \ddot{x} &= -2\Omega_b\dot{y}+\frac{\partial\Phi}{\partial x}+\Omega_b^2 x,
    \\
    \ddot{y} &= \phantom{+}2\Omega_b\dot{x}+\frac{\partial\Phi}{\partial y}+\Omega_b^2 y,
    \label{eqn:eomyd}
\end{align}
where x and y are Cartesian coordinates in the rotating frame,
$\Omega_b = 40$\ km\ s\inv\ kpc\inv is the pattern speed of the galactic bar, and $\Phi$ is the total potential with two components $\Phi = \Phi_{b}+\Phi_{d}$.

For the potential induced by the Galactic bar, the main contributor to non-axisymmetry, we adopt the Ferrers bar potential \citep{Ferrers1877QJPAM..14....1F, Perek1962AdA&A...1..165P, Vaucouleurs1972VA.....14..163D}. The Ferrers bar is an inhomogeneous spheroidal bar with a density distribution $\rho = \rho_0 (1-m^2)^2$, where $m^2 = x^2/a^2+(y^2+z^2)/c^2$. In this form, $c < a$ are the semi-major and semi-minor axis of the bar. The potential induced by such a bar is given by,

\begin{equation*}
    \Phi_b = \frac{105 G M_b}{32 \epsilon} \tilde{\phi}_b,
\end{equation*}
where $\epsilon^2 = a^2-c^2 = (5$\ kpc$)^2$ is a scale length determined by the semi-major ($a$) and semi-minor ($c$) axes of the bar, $M_b = 2 \times 10^{10}$\ $M_{\odot}$ is the mass of the bar, and $\tilde{\phi}_b$ can be obtained by recurrence relations. The expression of the potential can be found in closed form in \cite{Vaucouleurs1972VA.....14..163D} and Appendix \ref{app:fbar}. This model has been used by many studies regarding orbit families in barred galaxies \citep{Athanassoula1992MNRAS.259..328A, BinneyTremaine2008gady.book.....B, Kim2012ApJ...758...14K_usingferrers}. The Ferrers bar potential represents the potential induced by a prolate spheroid with parameters with physical meanings. Hence it is an idealised simple model of the bar but is sufficiently complicated to capture observational properties of the bar (e.g. the updated semi-major length $a \sim 5$\ kpc).

For the axisymmetric component, a simple 2D logarithmic potential is adopted to mimic a flat galactic rotation curve,

\begin{equation*}
    \Phi_d = -\frac{1}{2}V^2 \ln{(r^2+R_c^2)},
\end{equation*}
where $r^2 = x^2 + y^2$ is the radius away from the galactic centre. In this potential, $V$ represents the flat galaxy rotation curve in the galactic plane, and $R_c$ is a constant that controls the shape of the rotation curve in the inner Galaxy $R << R_c$. Parameters $V = 240$\ km\ s\inv and $R_c = 2.5$\ kpc are chosen to mimic the flat rotation curve found in many disc galaxies including the Milky Way \citep{Eilers2019ApJ...871..120E}.

We make this system dimensionless by normalising length and time. The lengths are normalised by the scale length, $r =\Tilde{r} \epsilon = \Tilde{r}\cdot(5\text{\ kpc})$, where $\Tilde{r}$ is the length in the system and $r$ represents the lengths in space with physical units. The times are normalised by the bar angular frequency, $t=\Tilde{t} / \Omega_b=\Tilde{t}/(40\text{\ km\ s\inv\ kpc\inv})$, where $\Tilde{t}$ is the time in the system, and $t$ represents the physical time.

Then, by dropping the tildes (writing $\Tilde{x}$ as $x$) and collecting constant terms, we obtain two dimensionless numbers,

\begin{align*}
    Q &= \frac{105 G M_b}{32 \Omega_b^2 \epsilon^3},\\
    P &= \frac{V^2}{\Omega_b^2\epsilon^2},
    \\
    \frac{Q}{P} &= 0.98(\frac{M_b}{2\times10^{10}\ M_{\odot}})(\frac{\epsilon}{5\ \text{kpc}})^{-1}(\frac{V}{240\ \text{km\ s\inv}})^{-2}.
\end{align*}

With these dimensionless numbers, the equations of motion in the dimensionless system can be written as,

\begin{align}
    \label{eqn:eomx}
    \ddot{x} &= -2\dot{y}+Q\frac{\partial\Phi_b}{\partial x} - P\frac{x}{r^2+R_c^2}+x,
    \\
    \ddot{y} &= \phantom{+}2\dot{x}+Q\frac{\partial\Phi_b}{\partial y} - P\frac{y}{r^2+R_c^2}+y.
    \label{eqn:eomy}
\end{align}

The modelled potential in the rotating frame, the effective potential, is presented in Fig.\ \ref{fig:potential}. The left panel shows a contour plot with the bar marked as a grey ellipse in the middle and the sun as the red asterisk. As shown, this barred potential has five stationary points, or Lagrange points, marked from L1 to L5. These points are stationary points in the effective potential where $\nabla \Phi_{\text{eff}}=0$, and static particles stay at these points in the rotating frame. Among the five points, L1 and L2 are non-stable saddle points, L3 is a stable minimum, and L4 and L5 are maxima, stable in most realistic potentials \citep{Sellwood1993RPPh...56..173S} and in our potential. Together, the effective potential shows a topology analogous to a ``volcano", as shown in the right panel of Fig.\ \ref{fig:potential}: the maxima L4, L5 and saddles points L1, L2 define the ``rim" together and the central minimum L3 sits in the ``crater" inside the rim. Under this analogy, the sun is located slightly outside the rim.

The rotation curve along the y-axis, generated by the separate and combined components of this model, is presented in the top panel of Fig.\ \ref{fig:rotcurve}. This potential, while simple, is sufficiently complex to capture stable in-plane features of the Milky Way, with a  long, slow bar and a flat rotation curve. The non-axisymmetric bar mainly contributes to the potential in the inner galaxy $R < 10$\ kpc. Outside $10$\ kpc, the rotation curve is dominated by the axisymmetric component: the disc and the dark matter halo. Although slightly higher, the combined rotation curve shows a good match to the observed rotation curve of the Milky Way \citep[e.g.][]{Eilers2019ApJ...871..120E}, which is marked by the dashed black line. The innermost galaxy ($R < 5$\ kpc) is dominated by the non-axisymmetric bar and is not well understood due to limits in observation and is therefore shaded.

The azimuthal force $\partial \Phi/(R\partial\phi)$ along the $25^\circ$ line passing our Sun and the Galactic centre is shown in the bottom panel. Since the logarithmic component of the potential is axisymmetric, the azimuthal forces are contributed only by the Ferrers bar. The bar-induced axi-asymmetry mainly exists in the innermost $\sim 6$\ kpc, and the galactic disc becomes nearly axisymmetric outside $7$\ kpc.  

\section{Analysis}\label{sec:dynamics}

To consider the Trojan scenario, we investigate the properties of Trojan orbits with the dynamical model in \S\ref{sec:orbits}. We then relate Trojan orbits to the kinematics observed in the SNd in \S\ref{sec:m2k}. Finally, we consider the stability of Trojan orbits with surfaces of section and look into other related orbit families in the model in \S\ref{sec:sos} and \S\ref{sec:y0ej}.

\subsection{Trojan orbits}\label{sec:orbits}

As a kinematic anomaly, the Hercules group is expected to have a dynamical origin. In this section, we adopt orbital dynamics to look into a popular scenario, in which the Hercules group is associated with the corotation resonance of the long slow bar \citep{PerezVillegas2017ApJ...840L...2P, Binney2020MNRAS.495..895B, DOnghia2020ApJ...890..117D}. We intend to understand the corotation origin at the orbit level and evaluate the likelihood of the Trojan scenario. 

Among the five Lagrange points, points on the bar minor axis, L4 and L5 are of primary interest. These points are stable maxima and can trap stars. These stars are called Trojans, and we call their orbits Trojan orbits. As static particles remain static at Lagrange points in the Corotating frame, the non-central Lagrange points are associated with corotation resonance.

Numerical integrations are conducted to obtain numerical solutions of stellar orbits in the 4D phase space. Thanks to the analytic potential in closed form, the potential gradients are analytically determined, and by setting up an initial condition in the phase space, the orbit can be simulated by numerically integrating the equations of motion (Equations \ref{eqn:eomx} and \ref{eqn:eomy}) in the rotating frame in the dimensionless system.

We limit our study to orbits that are symmetric about the y-axis (the bar minor axis). The initial positions of orbits are restricted to the y-axis and the initial velocities are set parallel with the x-axis and to the right $(x_0=0, \dot{y}_0=0, \dot{x}_0 > 0)$. This leaves $y_0$ and $\dot{x}_0$ as initial conditions, and as the direction of $\dot{x}_0$ is assumed, $\dot{x}_0$ can be replaced by the conserved quantity $E_J$. Because we adopt a time-independent potential in the rotating frame, a set of initial conditions determines an orbit. As numerical integrations in our work are well-conditioned, numerical solvers do not introduce further instability. Therefore, each set of ($E_J$, $y_0$) determines a specific orbit and we refer to the space of ($E_J$, $y_0$) as the initial condition space. In this study, the orbits are integrated for a time interval $\Delta t = 500$, which corresponds to about $12$\ Gyrs. The snapshots of the phase space ($x, y, \dot{x}, \dot{y}$) are recorded every 1/200 unit time (100,000 time steps in total). The numerical solutions are validated by the conservation of Jacobi integral $E_J$. In the study, $E_J$ remains constant to at least 6 significant figures.

Stellar motions in this system are non-linear and orbits can be divided into three categories: \textit{chaotic}, \textit{quasi-periodic}, and \textit{periodic} \citep{Sellwood1993RPPh...56..173S, BinneyTremaine2008gady.book.....B}. \textit{Chaotic} orbits only have one integral of motion $E_J$ and they are free to explore a hypervolume in the phase space $(\mathbf{r},\dot{\mathbf{r}})$. The randomness in these orbits suggests that they are more likely to form backgrounds than stable kinematic patterns. \textit{Periodic} or closed orbits, on the other hand, revisit the initial condition and form closed loops. The periodicity emphasizes the importance of these orbits due to their capability of forming and preserving stellar structures. However, the family of periodic orbits is a lower dimensional set in the initial condition space and thus has zero measure or probability. Hence, periodic orbits usually do not directly form structures. \textit{Quasi-periodic} orbits (QPO) oscillate around stable periodic orbits in a sense similar to the epicyclic motion in axisymmetric disc galaxies or stellar systems. Hence, these orbits conserve an integral in addition to $E_J$ and form 2D subsets of the initial condition space. These orbits cover a larger area of the initial condition space and form the backbone of the Trojan theory for the formation of the Hercules group.

\begin{figure*}
    \centering
    \includegraphics[width=\linewidth]{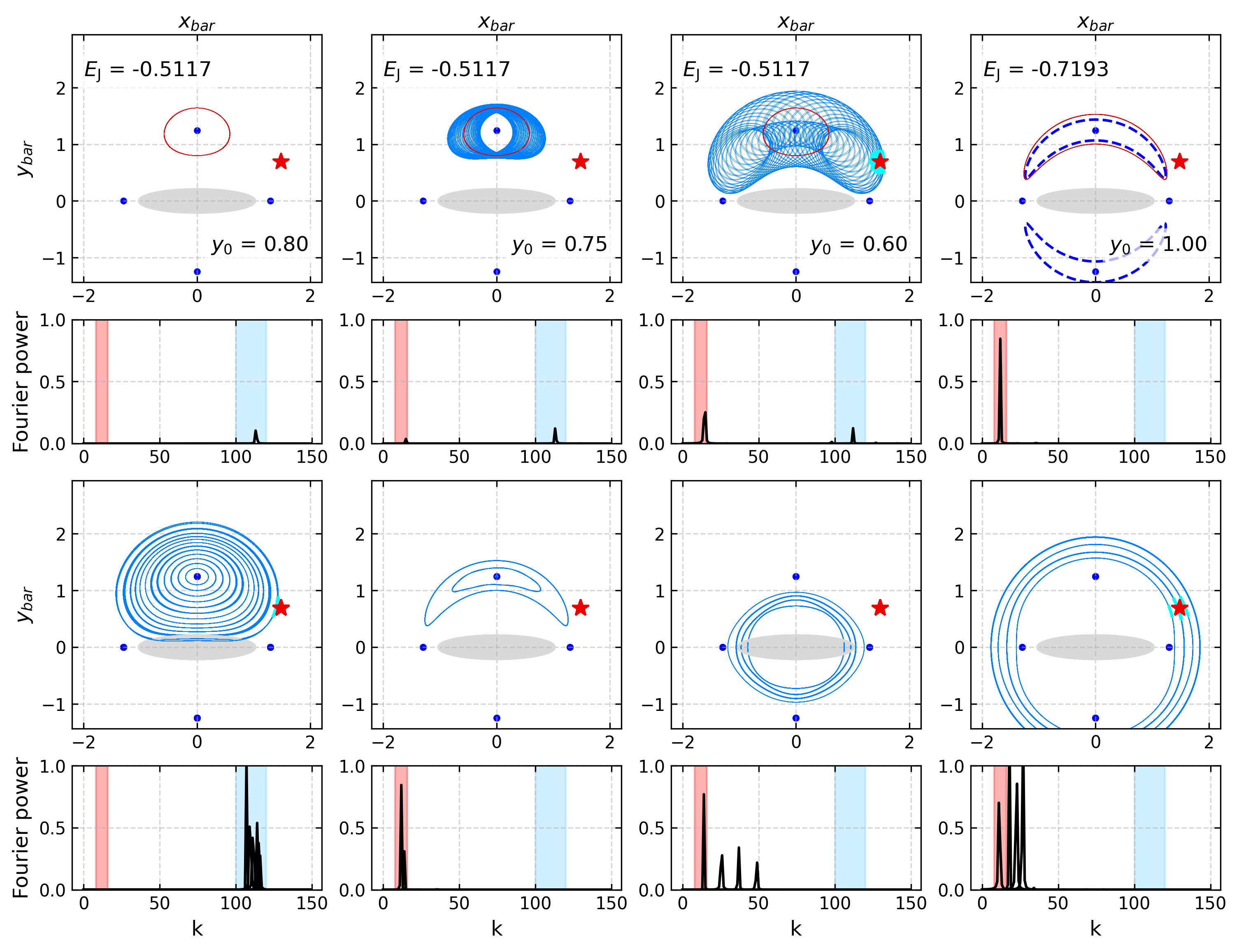}
    \caption{Families of periodic and quasi-periodic orbits and their corresponding Fourier power spectrum. The red asterisk marks the position of the sun and cyan marks the part of the orbit passing through the SNd. Top left: a fast Trojan periodic orbit; Top right: a slow Trojan periodic orbit that follows the ZVC marked by the dashed blue contour; Top middle: morphologies of two blue QPOs generated by perturbing the red fast periodic orbit on the top left; Bottom: four families of periodic orbits, from left to right: the more circular, fast Trojan family, the banana-shaped slow Trojan family, and the inner and outer circular families that follow the potential contours. The orbits in fast and slow Trojan families share similar frequencies in the power spectrum to other members of the family. The approximate range of the frequencies is shaded, slow family in red ($8<k<16$) and the fast family in blue ($100<k<120$).}
    \label{fig:orbits}
\end{figure*}

In our model, the minor axis Lagrange points L4 and L5 trap two families of periodic orbits, the slow, banana-shaped and the fast, more circular periodic orbits, presented in the bottom panel of Fig.\ \ref{fig:orbits}. Both families are retrograde, against bar rotation in the rotating frame. In physical units, the orbits in the slow Trojan orbits have frequencies about $f_s \sim 1$\ Gyr\inv while the fast Trojan orbits have frequencies about $f_f \sim 9$\ Gyr\inv. The slow orbits wallow through the potential as the Coriolis force balances the potential gradient, while the fast orbits are more energetic, cross through equipotentials, and exhibit much stronger Coriolis force to support the centripetal force required. As the effective potential is symmetric about the x-axis, all orbits on the near side of the galaxy exist centrosymetrically about the galactic centre on the other side. Hence, although we draw our attention primarily to the L4 orbits, we should note that centrosymmetric families also exist around L5.

These two periodic orbit families are both stable and can form families of quasi-periodic orbits under perturbations. Practically, orbits are determined by $E_J$ and $y_0$, and as $E_J$ is an integral that is conserved in an orbit, we achieve perturbations by slightly changing the initial position $y_0$ while holding $E_J$ constant. The quasi-periodic orbits can be found by perturbing in both directions. The top right plot in Fig.\ \ref{fig:orbits} shows a banana-shaped slow orbit, which follows the curves of zero velocity (ZVC). The ZVC occurs where the Jacobi integral of the orbit is below the maximum effective potential at L4, $E_J < max(\Phi_{\text{eff}}) = \Phi_{\text{eff, L4}}$. Therefore, from Equation \ref{eqn:ej}, there exists a region where the kinetic energy in the rotating frame becomes negative, and the region is non-explorable by the orbit. The QPOs of the slow family are more fragile under large perturbation while QPOs of the fast family are very stable, capable of tolerating large perturbations in the initial condition $y_0$. This results in very extended QPOs that travel as far as $5$\ kpc away from the parent orbit.

To better understand these quasi-periodic orbits as epicycles, we conduct spectral stellar dynamics analysis \citep{Binney1982ApJ...252..308B} on the orbits in the Trojan family. We Fourier transform the x-coordinates $x(t)$ of an orbit and plot the Fourier power spectrum against frequency under each orbit plot in Fig.\ \ref{fig:orbits}. On the top panel, we take a Trojan periodic orbit with $E_J = -0.5117$ and perturb it to show different morphological states of the QPOs. In each plot, the periodic orbits are coloured in red while the QPO perturbed from them is shown in blue. The leftmost orbit shows the parent periodic orbit with no perturbation applied. With very little perturbation, the near-circular orbit ``thickens" slightly and largely remains in the original shape. Under larger perturbations, the QPO forms very extended orbits that can reach the SNd. As we will see, these QPOs can form the building blocks of the Hercules structure. By considering the power spectra of two periodic orbits on the leftmost and rightmost plots in the top panel, we see that the circular fast periodic orbit corresponds to a single high frequency, with wave number $\kappa_f \sim 110$ and the banana-shaped periodic orbit corresponds to a single low frequency $\kappa_s \sim 12$. 

On the bottom panel, we present four periodic orbit families associated with the Trojan orbits and their corresponding single-peaked power spectra in the sample graph below. From left to right, we present the fast Trojan family, the slow Trojan family, the inner galactic circular family, and the outer galactic circular family. In the left two figures, we see that all fast periodic Trojan orbits share a similar frequency and all slow periodic Trojan orbits also share another frequency, both with a small spread. We then mark the two periodic frequencies in all plots with shaded semi-transparent blocks, the higher frequency in blue and the lower in red. Back to the upper panel, the fast and slow periodic orbits stay in their corresponding range of frequencies. On the other hand, each QPO in the middle has two frequencies: a high and a low frequency that falls into the frequency range of the fast and slow closed orbits. In the spectrum of the ``thickened" fast QPO, the fast frequency is the dominant frequency. The extended QPO shows an opposite combination, where the low frequency dominates the power spectrum, resulting in an overall banana-shaped morphology. Although not presented, the QPOs of the slow Trojan family only show the banana-shaped morphology and the Fourier spectra always show dominance in the slow frequency. We conclude that the QPOs of the Trojan families can be considered as the superposition of fast and slow modes of Trojan periodic orbits, and the morphology depends on the relative significance of the two frequencies. We note that the tiny side peaks around the fundamental fast and slow frequencies in the QPO plots are harmonics between the two frequencies.

We will see later in \S\ref{sec:y0ej} that the slow Trojan family bifurcates (breaks up) into an inner circular and an outer circular family of periodic orbits that also follows the ZVC. When the Jacobi integral becomes smaller than the effective potential at the saddle points L1 and L2, $E_J < \Phi_{\text{eff, L2}}$, the ZVC prohibits communication between the inner and outer galaxy, resulting in circular ZVC borders and the two circular families. In the lower, rightmost two plots of circular orbits, the frequency starts in the range of the slow frequency but moves to higher frequencies as the $E_J$ drops further and the orbits move away from the Trojan families.

Furthermore, the rim of the volcano in Fig.\ \ref{fig:potential} hinders most of the communication between the inner galaxy (inside the rim) and the outer disc (outside the rim). At typical $E_J$, the only y-symmetric orbits that explore both sides of the rim are the fast and slow Trojan families. Hence, they have the potential to transport inner galactic stars to the outer disc, and vice versa. As the Sun sits on the outside of the rim, if stars with inner Galactic origins can be found in the SNd, they are likely to be transported by Trojan orbits.

\subsection{Kinematics of Trojan orbits in the SNd}\label{sec:m2k}

\begin{figure*}
    \centering
    \includegraphics[width=\linewidth]{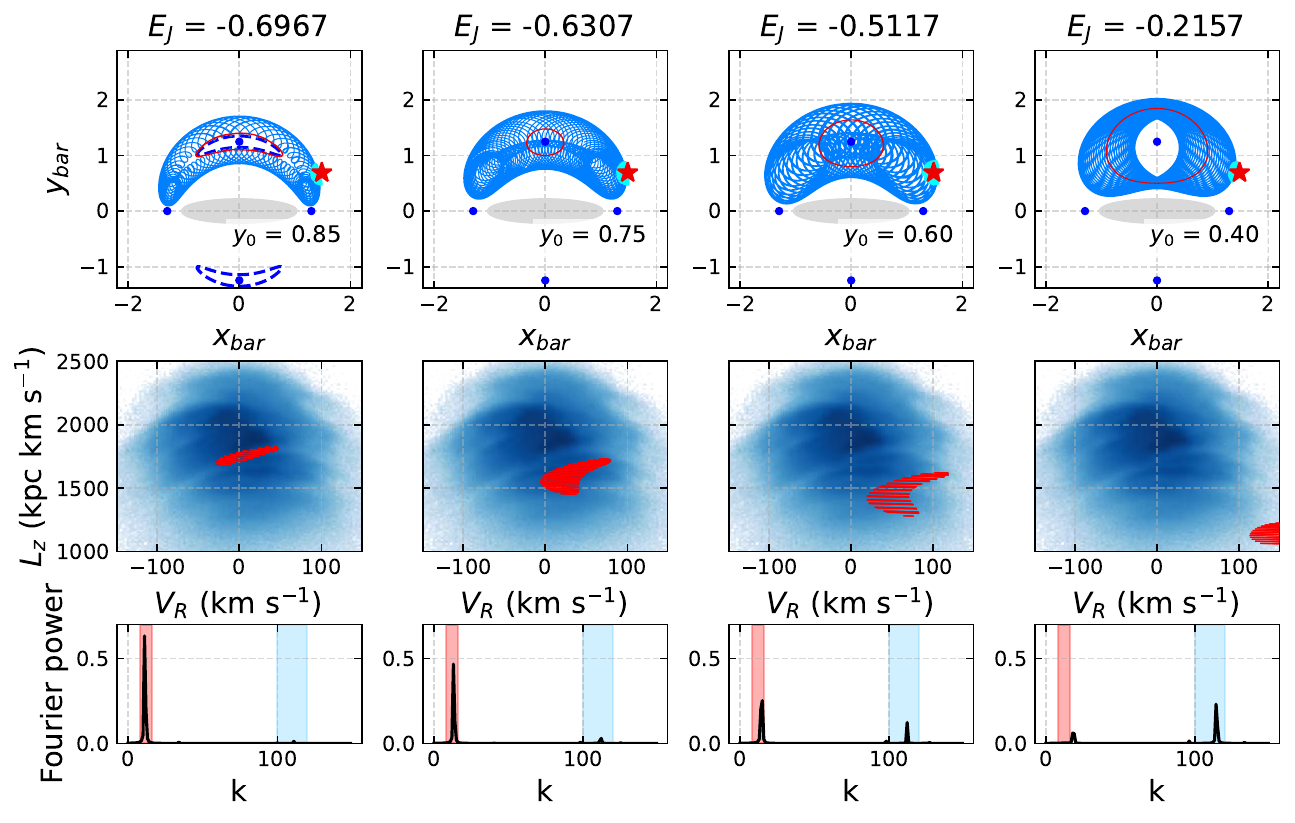}
    \caption{QPOs in the model mapped to the $L_Z$-$V_R$ kinematics plane by mock observation. \textit{Top row}: the morphology of blue QPOs perturbed from the red parent periodic orbits. The sun is marked as a red asterisk and the part of the orbits passing through the SNd are coloured cyan. The ZVC contours appear in the first row as it has an $E_J < \Phi_{\text{eff,}L_4}$; \textit{Middle row}: the $L_Z$-$V_R$ kinematics of the corresponding orbit when they pass through the SNd. The kinematics are coloured in red on top of the blue \textit{Gaia} kinematics distribution. The mapped kinematics covers the Hercules group and shows a bias towards positive $V_R$; \textit{Bottom row}: the Fourier power spectrum of the QPO in the same column. The shaded red and blue regions mark the frequency range of two periodic Trojan families.}
    \label{fig:m2k}
\end{figure*}

To associate the Trojan QPOs that pass through the SNd with the observed overdensities, we make mock observations by integrating the QPOs, obtaining their kinematical information when they pass through the SNd, and comparing their SNd kinematics to real data.

In the model, QPO families with a range of Jacobi integrals can reach the SNd, be observed, and get mapped into the kinematics plot. Fig.\ \ref{fig:m2k} shows a few of the most extended QPOs associated with each periodic orbit. The top panel shows the Trojan QPOs in blue and their parent periodic orbit in red. Their kinematics in the solar neighbourhood are drawn in the middle panel as red dots, on top of the blue kinematics distribution of \textit{Gaia} stars. Due to the finite time ($\sim$\ a Hubble time) of integration, these data appear as a set of stripes in the $L_Z$-$V_R$ plot, but the kinematics of the complete orbit would cover the whole area the stripes cover. Moreover, in reality, different stars can be found at different phases of the same orbit and cover the full area. We observe that over a range of Jacobi integral, when the QPOs reach the SNd, these orbits cover the Hercules structures and show the asymmetric structure in favour of positive $V_R$. 

Among the orbits, the least energetic orbit (with the lowest $E_J$) is perturbed from a slow periodic orbit. This QPO populates the highest angular momentum substructure Hercules I, and may possibly go even higher in $L_Z$ into the valley. Next to the right, a slightly more energetic orbit, perturbed from a fast periodic orbit populates Hercules II and a part of Hercules III. The next, more energetic orbit preserves the lowest angular momentum and covers part of Hercules III and Hercules IV. On the rightmost panel, a QPO with extremely high $E_J$ shows the morphology of the thickened circle. This orbit, while can still visit the SNd, includes quite extreme kinematics that get it mapped onto the edge of the plot.

When the perturbation is large, the kinematics map shows a region with a boomerang shape that is concave towards the bottom right. This curvature is similar to the observed curvature in the kinematics distribution constructed by the $V_R$-biased Hercules II, less-biased Hercules III, and a more biased Hercules IV. While the lower $E_J$ orbits cover regions highly populated by observed stellar kinematics, the highest $E_J$ orbit only visits an area in $L_Z$-$V_R$ with much lower densities. We attribute this to the fact that these orbits are very energetic, and hence less likely to exist. While they are capable of reaching the SNd to get mapped into the kinematics plot, the lack of members prevents them from forming patterns observable in the kinematics histogram. 

The bottom panel of Fig.\ \ref{fig:m2k} shows the Fourier-transform power spectrum of the corresponding QPOs. The QPO perturbed from the slow, banana-shaped periodic orbit has the majority of its power in lower frequency, with very little power in the fast frequency. The principal frequency of this QPO with the slow parent orbits is always the slow family. By perturbing the fast family, the parent periodic orbit first ``thickens" and then forms the extended morphology. The fast family with the lowest $E_J$ holds QPOs with the thickened morphology that transforms into the extended morphology easily under larger perturbations. With the increase of $E_J$, the thickened morphology become more stable relative to the extended morphology. While the extended morphology can survive under large perturbations, they are close to the limit of stability and breaks into inner disc circular families further away.

\begin{figure*}
    \centering
    \includegraphics[height=.85\textheight]{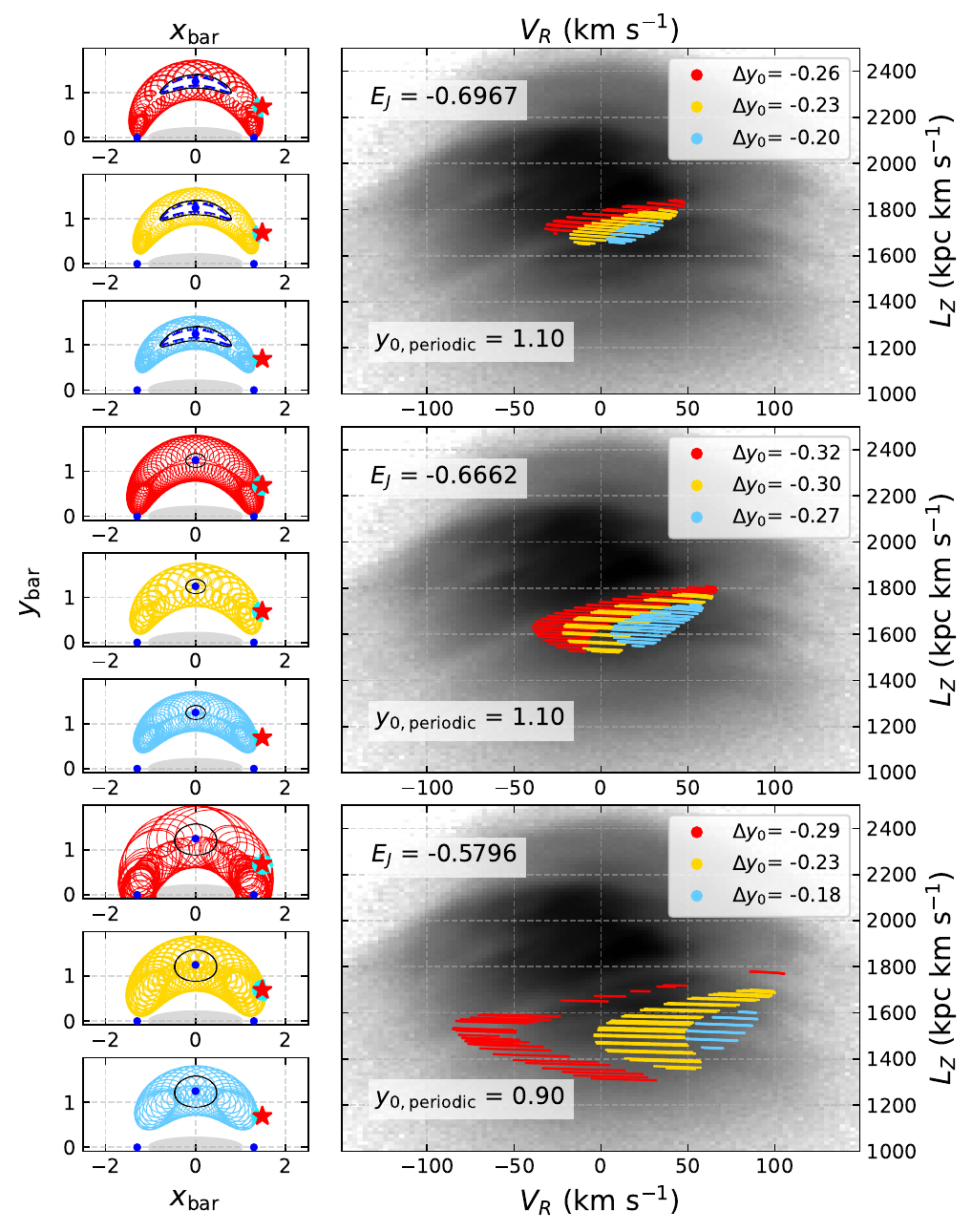}
    \caption{The variation of mock observations of the QPOs in the kinematics space with $\Delta y_0 = y_{0,\text{QPO}} - y_{0,\text{periodic}}$. The kinematics of QPOs in the SNd with three different values for $E_J$s are presented on the right. For each $E_J$, three QPOs with different perturbations are presented. The morphologies of the corresponding orbits are plotted on the left in the corresponding colour. From bottom to top in each panel, the perturbation increases from the edge of reaching the SNd to the edge of breaking. The stripes in the kinematics plot are due to the finite numerical integration time. The complete orbit covers the whole area the stripes cover. Together, the Trojan orbits at different $E_J$ and $y_0$ cover the Hercules group and exhibit the triangle shape that agrees with the exterior of the density distribution.}
    \label{fig:m2k_y0}
\end{figure*}

We also investigate how the SNd kinematics vary in the initial condition space ($y_0$, $E_J$). In Fig.\ \ref{fig:m2k_y0}, we hold $E_J$ constant and vary the initial position $y_0$ in the range of the SNd reaching orbits. For each $E_J$, we plot the morphology of three QPOs on the left, differentiated by different colours. Then, their SNd $L_Z$-$V_R$ kinematics are plotted on the right, labelled by the difference in $y_0$ to the parent periodic orbit $\Delta y_0 = y_{0,\text{QPO}}-y_{0,\text{PO}}$. The parent periodic orbit of each QPO is presented in black. 

The top plot corresponds to the $E_J$ value of a slow family while the middle and bottom plots correspond to two fast families. The figures show that the QPOs cover larger spaces and higher $L_Z$ when perturbed further away from the periodic orbit. The higher $E_J$ orbit can hold QPOs with larger perturbations and cover a wider space. Starting from the closest orbit in blue, the QPO is mapped to a small egg at first, then distorts towards a boomerang. Larger perturbation increases the distortion of the shape of the region, resulting in larger boomerangs. Moreover, QPOs with lower $E_J$ move further up in $L_Z$ as the perturbation increases, while QPOs with higher $E_J$ are shifted left in $V_R$. In the presented figures, the QPOs with the lowest $E_J$ cover both Hercules I and II, and the higher $E_J$ orbits cover Hercules II, III, and a portion of Hercules IV.

We note that, at the lowest $E_J$, slow family QPOs have such high $L_Z$ that their SNd kinematics go over the valley and join the top fraction of the plot. Several potential explanations can be considered. As noted in \S\ref{sec:model}, our model has a rotation curve slightly higher than the observed Galactic rotation curve. This may result in a systematically higher $L_Z$ in orbits. Another option is to allow Hercules orbits to enter part of the Hyades group. While we observe the existence of a ``valley" marked by the non-smooth outskirts between two fractions, a significant fraction of stars still exist around $|V_R| \sim 0$. In this scenario, the more $V_R$ narrowed low $E_J$ QPOs naturally shrink the extended $V_R$ coverage in Hercules III and II. Indeed, some orbits with initial conditions close to Hercules Trojans are found to preserve $L_Z$-$V_R$ kinematics in agreement with the Hyades group. Other orbits with LSR and Arch/hat kinematics are also observed. These orbits may be potential candidates to help explain the origin of other kinematic groups (see \S\ref{sec:discussion}).

Overall, we conclude that, among QPOs observed in the SNd, orbits with higher $E_J$ tend to preserve lower $L_Z$. QPOs further away from their parent periodic orbits include kinematics with higher $L_Z$ and cover more distorted regions. The lower $E_J$ Trojan QPOs cover Hercules structures with higher $L_Z$ and the higher $E_J$ Trojan QPOs contribute more towards the lower $L_Z$ Hercules subgroups. We note that these results are sensitive to bar parameters, expecially the pattern speed $\Omega_b$. However, the current model $\Omega_b \sim 40\text{\ km\ s\inv\ kpc\inv}$ is strongly supported \citep{Sormani2015MNRAS.454.1818S, Portail2017MNRAS.465.1621P, Monari2019A&A...626A..41M, Drimmel2023A&A...670A..10D, Leung2023MNRAS.519..948L}.

\subsection{Surfaces of section}\label{sec:sos}

To understand the relation between the QPOs and periodic orbits, and the stability of orbits in our model better, the Poincar\'e surfaces of section are generated for two $E_J$s, one hosting a fast periodic Trojan orbit, the other hosting a slow rotating Trojan periodic orbit. In a two-dimensional planar system, the information of orbits is fully embodied in the four-dimensional phase space $(x,y,\dot{x},\dot{y})$. By using the conservation of $E_J$, the dimension can be reduced to 3D without loss of information. Then by taking a slice through the plane of symmetry, orbits can be distinguished by the topology in the plane, as a reflection of features in the phase space \citep{Henon1964AJ.....69...73H, Binney1985MNRAS.215...59B}. 

Chaotic orbits, due to their freedom to explore a higher dimensional region of phase space, cover a 2D area in the surface of section when integrated infinitely. Often, regions explored by chaotic orbits overlap with other chaotic orbits and show no definite pattern on the surface. Periodic orbits, on the other hand, form closed loops in phase space with one or higher multiplicity, and hence, when cut by a plane, they appear as single or a finite set of 0D points. Around loops of stable periodic orbits, the quasiperiodic orbits are confined to, yet free to explore the surface of the doughnut-shaped tori surrounding the parent periodic orbits. Hence, these orbits appear as one or a set of closed, sometimes very elongated and twisted, 1D curves centred at points representing their parent orbits, the invariant curves. The extent of perturbation the child QPOs can sustain before evolving into chaos or another family works as a measure of the stability of the periodic orbit.

In this study, the surfaces of section of symmetric orbits are generated by cutting through $\dot{y}=0$ while ensuring $\dot{x} > 0$ at a constant Jacobi Integral $E_J$. This generates a surface of section on the plane $(x,y)$. Other works (e.g \citealp{Athanassoula1983A&A...127..349A, Sellwood1993RPPh...56..173S, Cincotta2000A&AS..147..205C}) often generate surfaces of section in a position-velocity plane, e.g. ($y$, $\dot{y}$) or ($y$, $p_y$). The choice of the surface is arbitrary and does not affect the conclusions. Our choice facilitates the relation between spatial features of the orbit and the phase space features on the surface of section.

\subsubsection{Surface of section in $E_J = -0.6662$} \label{sec:sos_high}

\begin{figure*}
    \centering
    \includegraphics[width=\linewidth]{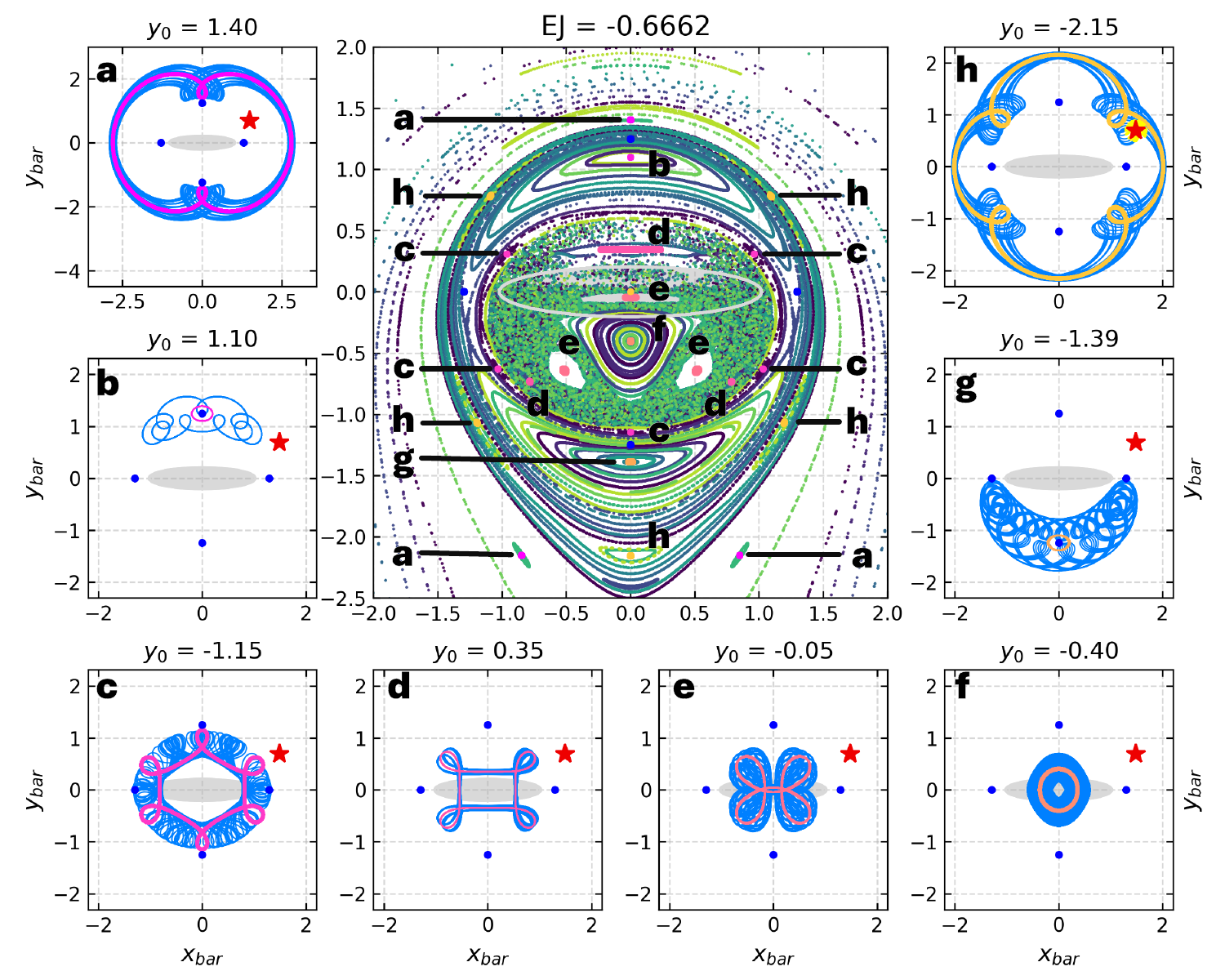}
    \caption{The $x$-$y$ surface of section and the morphologies of symmetric periodic orbits at $E_J = -0.6662$. In each morphology plot a to h, the periodic orbit is plotted in a warm colour on top of a blue QPO. The sun is marked as a red asterisk and the orbits passed through the SNd are highlighted in yellow. The corresponding points are marked in the same colour and with the label of the plot on the surface of section in the main panel. The Galactic bar and Lagrange points are marked the same as in the previous plots. The Trojan orbit orbits (b and g) are found central symmetrically around both L4 and L5. As indicated by its coverage in the surface of section, it is a highly stable orbit family.}
    \label{fig:sos_high}
\end{figure*}

First, we present the surface of section of an $E_J > \Phi_{\text{eff, L4}}$, which hosts a fast Trojan family. We generate the surface of section plot of $E_J = -0.6662$ with equally spaced initial positions from $y_0=-2.5$ to $y_0=2.5$ with steps of 0.05. The plot of the surface of section is present in the middle of Fig.\ \ref{fig:sos_high}. The morphology of periodic orbits in the ($x$, $y$) plane is shown in subplots \textbf{a} to \textbf{h}. In each morphology plot, a quasi-periodic orbit is plotted in blue under their respective parent periodic orbit. The points in the surface of section plot that correspond to the periodic orbits of each subplot are drawn in the same colour and labelled. As in previous figures, the position of the sun, the Lagrange points, and the bar are marked respectively. Paths of the QPOs are coloured yellow while passing through the solar neighbourhood.

\begin{itemize}
    \item At the top left, with the highest $y_0$, in Fig.\ \ref{fig:sos_high}a, we see an outer disc orbit that turns twice at the bar minor axis outside the L4 and L5 Lagrange points. This is the -2:1 resonance orbit, which is associated with the OLR \citep{Fragkoudi2019MNRAS.488.3324F, Asano2020MNRAS.499.2416A}. This orbit has two turning circles on the bar minor axis towards the galactic centre and, in the surface of section, is responsible for the three-point structure at about $(\text{xs, ys}) = (0,1.4)$ and $(\pm 0.8,-2.2)$, symmetric about the y-axis in the surface of section. Orbits like this are believed to be responsible for the Hercules structure in the fast bar scenario but are beyond the solar neighbourhood in this slow bar model.
    
    \item The series of concentric banana shapes centred at $(0,1.1)$ in the surface of section corresponding to the fast rotating Trojan periodic orbit in Fig.\ \ref{fig:sos_high}b. When perturbed up and down, thickened and extended QPOs form and appear as the thin-elliptical and banana-shaped invariant curves surrounding the point representing the periodic orbit on the surface of section. Before breaking into chaotic orbits, the Trojan family forms a -7:1 island periodic orbit. As seen in subplot \ref{fig:sos_high}b, this orbit shows seven turning circles towards the L4 point, in a similar sense to outer galactic resonance orbits towards the galactic centre. These orbits form when two frequencies in the quasi-periodic orbit form a rational ratio.
    
    \item Towards the inner galaxy, a 6-turn inner disc hexagonal orbit is found in Fig.\ \ref{fig:sos_high}c. While the specific periodic orbit is generated at $y_0 < 0$, its QPOs are also found around $(0,0.6)$, below the L4 Trojans in the surface of section. This orbit shows the morphology of a hexagon with turning circles away from the galactic centre. In the surface of section, it is responsible for the five-point structure surrounding the inner galactic disc, at $(0,-1.2)$, $(\pm 1,-0.6)$, and $(\pm 0.9,0.3)$. This family explores two sides of the inner galaxy and is less stable.

    \item Further down, the family breaks into irregular inner galactic chaotic orbits. As seen in the surface of section, this chaotic region covers most of the inner galactic phase space except for three regular regions. Among the three, the only orbit family with a positive $y_0$ is the rectangular orbits in Fig.\ \ref{fig:sos_high}d. This is a 4-turn family that follows the exterior of the galactic bar. This family forms the three-point structure at $(0,0.4)$ and $(\pm 0.8,-0.7)$ in the surface of section. This orbit belongs to the x1 orbit family that is believed to be the building block of the galactic bar \citep{Athanassoula1983A&A...127..349A, Athanassoula2009MNRAS.400.1706A}. Unlike the Trojan family, the QPOs of the rectangular family do not change in morphology and only have thickened QPOs. The family breaks into chaos further in.

    \item Into the centre of the galaxy in the surface of section plot, we find another 4-turn family that has the morphology of a butterfly in Fig.\ \ref{fig:sos_high}e. This orbit also induces a three-point structure in the surface of section at $(0,0.1)$ and $(\pm 0.5, -0.7)$. We note that the stability of this orbit family can be affected by the existence of the central supermassive black hole and its nuclear stellar disc, so the periodic orbit may not appear in more complicated models like \cite{Hunter2024arXiv240318000H}. 
    
    \item In Fig.\ \ref{fig:sos_high}f, we see a very stable and extended circular family at $(0,-0.4)$, retrograde in the rotating frame. This orbit has a very high frequency. The QPOs of this family thicken the circle by the typical rosette shape, as the superposition of two frequencies. This family is the most stable and important orbit family in the inner galaxy. As we see in Appendix \ref{app:sos_evo}, this family exist throughout all the $E_J$s explored and dominates the phase space as $E_J$ increases.
    
    \item Below the L5 Lagrange point, we find the Trojan family on the other side of the galaxy in Fig.\ \ref{fig:sos_high}g. Due to the symmetry of the effective potential along the bar major and minor axes, this family shares the same morphological features as the L4 Trojan family. By further perturbing the orbit to the bottom, we find a circular QPO that explores both sides of the outer disc plane, presented in the third row of Fig.\ \ref{fig:futureorbits}. While we do not find a parent periodic orbit associated, this orbit may be associated with other high $L_Z$ kinematic groups like the Hyades group.
    
    \item The periodic orbit with the lowest $E_J$ is found in Fig.\ \ref{fig:sos_high}h. It is a 4-turn outer disc orbit that shows a -4:1 resonance. It appears as the five-point structure at $(0,-2.15)$, $(\pm -1.2, -1.1)$, and $(\pm 1, 0.7)$. This family has QPOs in similar morphology to that of the -2:1 family, and some have argued a possible scenario in which these orbits would contribute to the Hercules structure \citep{Hunt2018MNRAS.477.3945H_OLR, Asano2020MNRAS.499.2416A}. 
\end{itemize}

\subsubsection{Surface of section in $E_J = -0.7193$} \label{sec:sos_low}

\begin{figure*}
    \centering
    \includegraphics[width=\linewidth]{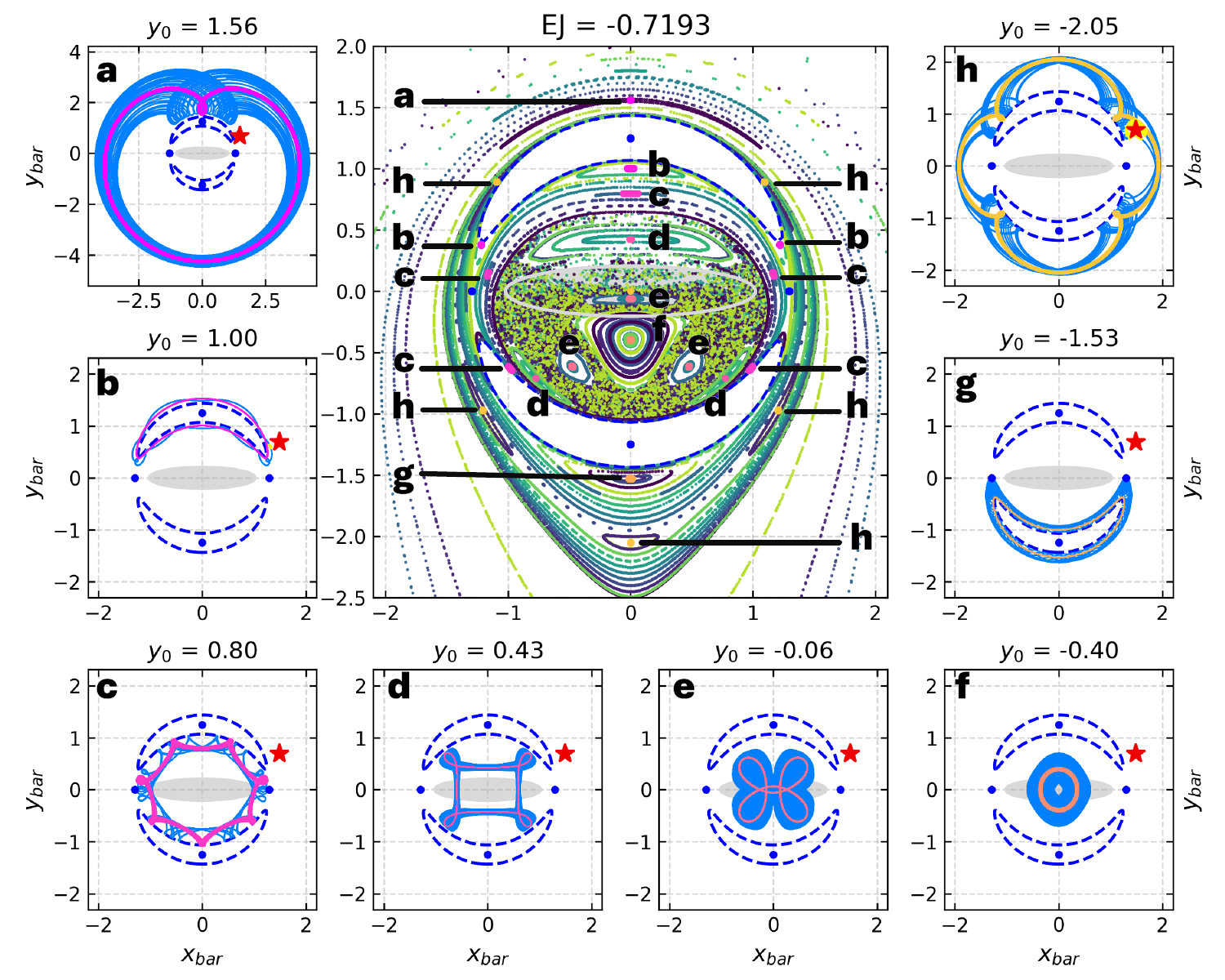}
    \caption{The $x$-$y$ surface of section and the morphologies of symmetric periodic orbits at $E_J = -0.7193$. The ZVC is marked by the two banana shaped blue dashed contours. Same as Fig. \ref{fig:sos_high} except for a different $E_J$. The slow Trojan family is more restricted by the ZVC. It covers a much smaller part of the surface of section, and is less stable.}
    \label{fig:sos_low}
\end{figure*}

In this section, we present the surface of section with an $E_J$ that includes a slow Trojan family. We investigate the impact of lower energy and the existence of ZVC on the phase space of orbits. In Fig.\ \ref{fig:sos_low}, the surface of section plot of $E_J = -0.7193$ is generated in the middle panel with periodic orbits and their corresponding QPOs on the sides, in a similar way to that in \S\ref{sec:sos_high}. 

\begin{itemize}
    \item Above the ZVC, we find a very large -1:1 resonance orbit in Fig.\ \ref{fig:sos_low}a. Different from the -2:1 orbit in Fig.\ \ref{fig:sos_high}a, this orbit only turns once above the L4 Lagrange point. Unlike most orbits on the high $E_J$ surface, this orbit does not show symmetry about the bar's major axis. In the surface of section, it corresponds to the point at $(0,1.6)$ and $(0,-4)$, which is outside the presented surface. 

    \item In Fig.\ \ref{fig:sos_low}b, we present the banana-shaped slow Trojan orbit with the -7:1 island orbit around it. On the surface of section plot, it appears as three points at $(0,1)$ and $(\pm 1.3,0.4)$, in the middle and two edge points of the ZVC. The island orbit at this energy shows much smaller turning circles than that at the higher $E_J$. 

    \item Further below, the family of bar surrounding orbits stabilises and shows a 7-turn, 7:1 resonance orbit at the place where the hexagonal orbit is located in Fig.\ \ref{fig:sos_high}. While with a different multiplicity, it also shows a five-point structure in the surface of section at $(0,0.8)$, $(\pm 1,-0.6)$, and $(\pm 1.2, 0.1)$. Similar to the 7-turn island orbit in subplot b, the turning circles are also significantly smaller than that in the hexagonal orbit.
    
    \item In Fig.\ \ref{fig:sos_low}d we encounter again the rectangular family. This rectangular orbit is located at $(0,0.4)$, and $(\pm 0.8,-0.7)$ in the main panel. The quasi-periodic orbits of this family cover a larger area of the surface of section, showing an increase in significance and stability in the phase space. The turning circles of this orbit also shrink.

    \item The butterfly orbits in the galactic centre and the inner galactic circular orbits undergo very little morphology transition, except with smaller turning circles and size. The butterfly orbit in subplot \ref{fig:sos_low}e is mapped to $(0,0)$ and $(\pm 0.5,-0.6)$ in the surface of section, and the circular orbit in \ref{fig:sos_low}f is located at $(0,-0.4)$. The two inner galactic orbit families both shrink in the surface of section coverage space as they are limited more by the surface of section.

    \item On the other side of the ZVC, the L5 slow Trojan orbit is found symmetrical to the L4 family and is mapped onto the surface of section at $(0,-1.5)$. The likely Hyades-related circular QPO is also found by further perturbing the slow Trojans. However, unlike the QPO in \S\ref{sec:sos_high}, although with a different morphology, this QPO is found to be a child of the slow Trojan family. See \S\ref{sec:discussion} for a more detailed idea.

    \item Further below, we arrive at the -4:1 resonance orbit that corresponds to the five-point structure at $(0,-2)$, $(\pm 1.3,-1)$, and $(\pm 1.2, 0.9)$ in the main panel. While morphologically analogous to that in \S\ref{sec:sos_high}, the size of turning circles is also reduced.
\end{itemize}

In general, most orbit families in \S\ref{sec:sos_high} continue to lower $E_J$. All turning circles of resonance orbits show a size reduction regardless of whether they are directed toward the galactic centre, away from it, or toward a Lagrange point (as in the case of the -7:1 Trojan island). The size of the turning circles is expected to continue to decrease as the $E_J$ goes down, which ultimately results in a change in orbit morphology or encounters a change in the morphology of the phase space. Some $n:1$ orbits go through morphological changes in the multiplicity of the resonance with the shift in $E_J$. Along with the coverage of their quasi-periodic orbits, this indicates a lack of stability in these orbit families. Some additional materials on orbits and the evolution of surface of section over $E_J$ can be found in Appendix \ref{app:sos_evo}. Overall, surface of section plots in Appendix \ref{app:sos_evo} emphasizes the Trojan family and the inner galactic retrograde circular family as the most stable orbit families throughout $E_J$ in addition to the ordinary circular disc families.

Back to the Trojan theory, the fast Trojan family covers a significant portion of the surface of section. While more restricted by the banana-shaped ZVCs, the slow Trojan family also covers a space with a decent amount of surrounding QPOs. These behaviours indicate the strong stability of the Trojan family and the high significance in the phase space. Therefore, we can expect a significant amount of stars to be trapped in this Trojan family. With a high distribution of the stars in the stable Trojan orbits, it can construct the over-density identified as the Hercules group in Fig.\ \ref{fig:gaialzvr}.    

\subsection{Trojan families in the initial condition space}
\label{sec:y0ej}

\begin{figure}
    \centering
    \includegraphics[width=\linewidth]{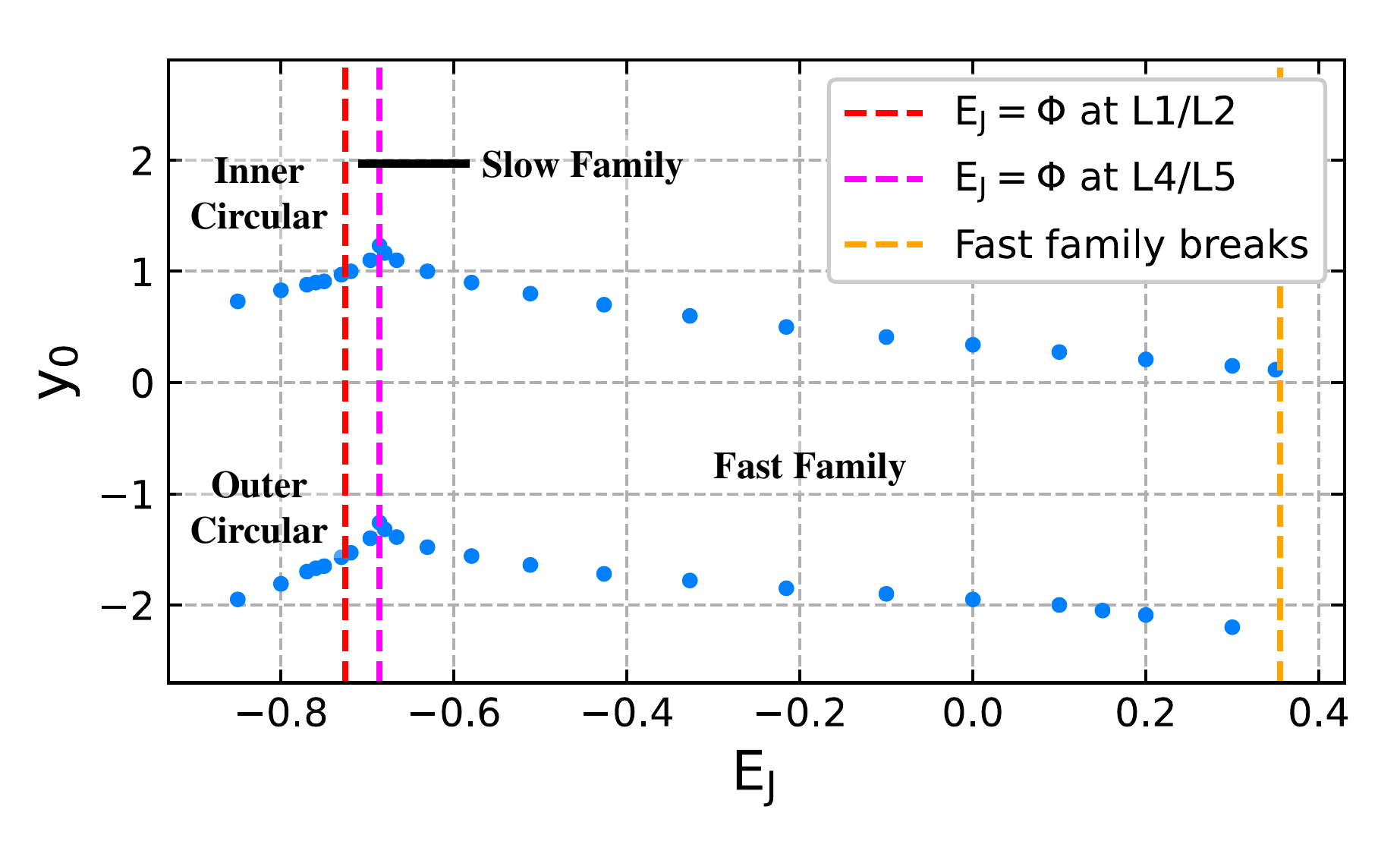}
    \caption{The $(E_J,y_0)$ initial condition space of the Trojan and related periodic orbits. $E_J$ is the conserved Jacobi integral and $y_0$ is where the orbit crosses the y-axis with $\dot{y}=0$ and $\dot{x} > 0$. The blue points mark samples of periodic orbits. The vertical dashed orange line marks the maximum $E_J$ that allows the existence of stable fast Trojan periodic orbits. The dashed magenta line marks the $E_J = \Phi_{\text{eff}}$ of the L4 and L5 Lagrange points. The dashed red line marks the $E_J = \Phi_{\text{eff}}$ of the L1 and L2 Lagrange points. The fast family transitions into the slow family between the red and magenta lines. The slow family breaks up into an inner galactic circular family and an outer galactic circular family left to the red line.}
    \label{fig:y0Ej}
\end{figure}

In addition to recognising the importance of Trojan orbits in the phase space with the method of surface of section, it is also curious to learn the stability of the periodic Trojan orbits in the $(E_J,y_0)$ initial condition space.

In Fig.\ \ref{fig:y0Ej}, we plot samples of the two Trojan families of periodic orbits and the related inner and outer circular families in the initial condition space. The $\Phi_{\text{eff}}$ of the Lagrange points are marked by dashed lines. Among the four-orbit families, the fast Trojan family dominates the initial condition space. Stable fast Trojan periodic orbits are found for $ -0.686 < E_J < 0.355 $. The fast family becomes unstable and breaks up at $0.355 < E_J $. Note that for each $E_J$, two fast Trojan periodic orbits are found, the top with positive $y_0$ around the L4 and the bottom with negative $y_0$ around L5. At the magenta line, we have $E_J = \Phi_{\text{eff, L4}}$ and the near-circular fast Trojan orbit reduces to a stationary particle at L4/L5. Together with L1 and L2 these are the points on which stationary particles sits and corotates with the bar, i.e. corotation. 

To the left of the magenta line, the Jacobi integral is below the maximum effective potential at L4/L5. Thus, ZVC occurs and the fast Trojan orbits transit into a family of slower orbits following the boundaries of the ZVC. Between the red and magenta line, $\Phi_{\text{eff, L1}} < E_J < \Phi_{\text{eff, L4}}$, we observe a small range of $-0.726 < E_J < -0.686$ where the ZVC shows a banana-shaped morphology. With the $E_J$ in this range, particles are not allowed to overcome the potential of the maxima points L4 and L5, but can still find paths through the saddle points L1 and L2 to communicate between the inner and outer disc. By following the ZVC boundaries, the particles form the slow Trojan family and show a similar banana-shaped morphology. 

At lower energy, to the left of the red line, the $E_J$ is lower than the effective potential of saddle points L1 and L2. With $E_J < -0.726$, the banana-shaped ZVCs at both sides of the galaxy join at L1 and L2. The new ZVC forbids communication between inner and outer discs and forms two near-circular boundaries. Then, the L4 slow Trojan orbit is encircled by the inner boundary and evolves into an inner disc circular family, prograde in the rotation frame, while the L5 orbit lies outside the outer boundary and evolves into a retrograde outer disc circular family. The morphologies of the two families are shown in the bottom right panels of Fig.\ \ref{fig:orbits}. As the outer family is less impacted by the asymmetry of the bar, it is expected to be similar to the classical circular orbits in axisymmetric models while the inner family is still affected by the asymmetry induced by the bar. The two families become two of the most stable and significant families in lower energies as seen in Appendix \ref{app:sos_evo}.

\section{Discussion} \label{sec:discussion}

In \citetalias{LYS2024arXiv241119085L:I}, we discovered a chemically distinct stellar population in Hercules III and IV. These stars are likely to originate from the outer Galactic bar. In our QPO Trojan orbit scenario, these stars can be born on or get caught by the L4 Trojan quasi-periodic orbits which take them through the solar neighbourhood. As more energetic QPOs with higher $E_J$ can cover a larger range of radius in the disc and include lower angular momentum while passing through the SNd, a larger portion of low $L_Z$ Hercules subgroup stars is likely to be formed in the inner Galaxy, get carried out to the SNd by the high $E_J$ Trojan orbits, and result in a patch of more enhanced [Fe/H] with low $L_Z$ at Hercules III and IV. On the other hand, as Hercules subgroups with higher $L_Z$ are more likely to have lower $E_J$, which corresponds to less radially extended QPOs, Hercules I and II are more LSR-like in chemical abundances.

Moreover, considering that the sun is located slightly outside the rim of the potential volcano in Fig.\ \ref{fig:potential}, transporting stars out from the inner Galaxy requires the underlying mechanism to be capable of overcoming the potential rim. As seen in \S\ref{sec:sos}, most stellar orbits in the galactic plane are limited by the potential rim and stay either inside or outside the rim at the presented $E_J$s. The fast and slow L4 Trojan orbits appear to be the only orbits at these $E_J$ that can cross the rim and thus be capable of transporting inner Galactic stars out. At lower $E_J$, as discussed in \S\ref{sec:y0ej}, the ZVC prohibits the communication between the inner and outer disc, and hence no orbit can cross the rim. At higher $E_J$, the turning circles expand with the increase of $E_J$ and can result in rim-crossing orbits. However, they generally do not include the right kinematics of Hercules and are more likely to contribute to the background. In addition, higher $E_J$ requires higher energy, resulting in a lower probability: while these orbits are stable and capable of transporting, they are less likely to exist in the disc. Therefore, the Trojan orbits present a scenario that can bring inner Galactic stars out in an ordered way with the kinematics of Hercules. This feature of Trojans is consistent with the chemical signatures of the observed Hercules kinematic group.

Furthermore, the surfaces of section in Appendix \ref{app:sos_evo} reveals the Trojan families, especially the fast family, as the second-most stable family throughout $E_J$. The most stable family is the retrograde inner galactic circular orbit shown in Fig.\ \ref{fig:sos_high}. These orbits are inner galactic orbits that cannot visit the SNd even at extremely high $E_J \sim 0$. Therefore, considering the stability and the capability to transport inner disc stars out, the Trojan orbits are the most likely orbits that could be responsible for the Hercules group in the SNd.

Apart from the inner disc stars born on the Trojan orbits, we are curious about how stars in the inner disc can be captured into the Trojan orbits and move outwards to the SNd. When the stars are close to the bar, the asymmetry of the bar potential defines the Jacobi integral $E_J = E - \Omega_b \cdot L_Z$ as the only general integral of motion. Therefore, unlike symmetric disc stars that have the conservation of energy and stay at a point in the $(L_Z, E)$ plane, bar-related orbits move on the straight line with a constant gradient of $\Omega_b$ and different offsets defined by the constant $E_J$. Then, if an unstable orbit approaches a Trojan orbit that happens to be on the same line in the $(L_Z, E)$ plane, it may be captured into the more stable Trojan orbit. Consider that the chemical abundances suggest a Hercules origin close to the super-thin-outer-bar, while the exact mechanism supporting the super-thin-bar is not yet well understood, its thickness creates a plausible scenario in which orbits that are unstable under vertical perturbations support the unusually thin structure. Then, if an unstable super-thin-bar orbit can be found fulfilling the previous conditions, it may be captured by a Trojan orbit and escape the potential of the bar. Another region close to the outer bar is the L1 and L2 Lagrange points. As unstable critical points, while orbits around them are unstable in nature, if certain orbits or gas can be trapped around, the instability should allow them to be captured easily by other orbits, including the Trojans.

To consider these questions, we need a less general, more realistic, and more complicated self-consistent model of the Galaxy to consider orbits in the inner Galaxy and the origin of the super thin bar. We plan to investigate these issues in the future. We plan to adopt a more realistic, observation-based, 3D potential of the Milky Way that includes details of the inner Galaxy with a boxy peanut-shaped bulge and a super thin long bar \citep{Sormani2022MNRAS.512.1857S, Hunter2024arXiv240318000H}. We will use this more accurate model to consider possible inner galactic orbits that might be captured by the Trojan orbits and the origins of the super thin bar that has a vertical scale height of only 40\ pc \citep{Wegg2015MNRAS.450.4050W}.

\begin{figure*}
    \centering
    \includegraphics[width=.9\linewidth]{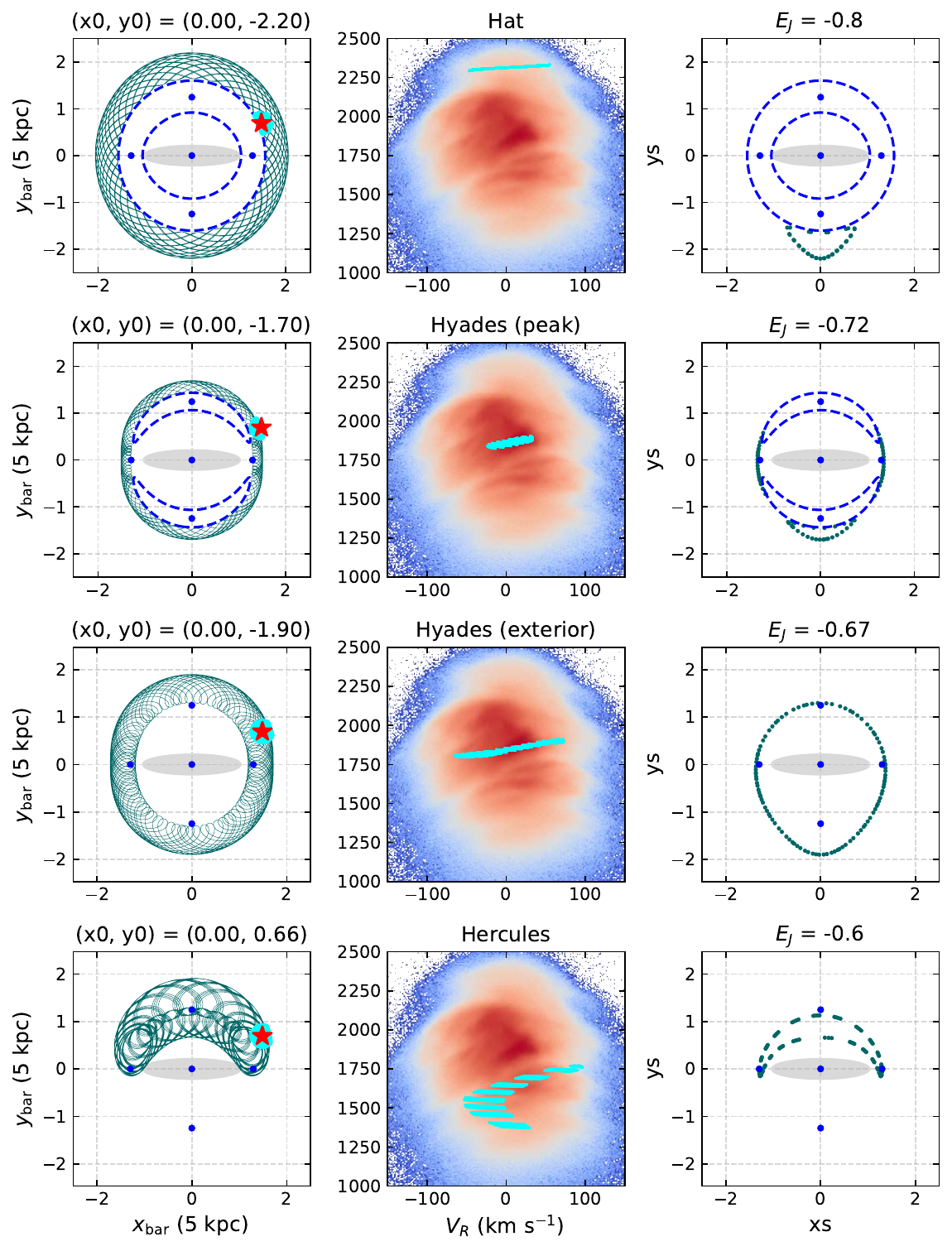}
    \caption{Examples of orbits that can potentially contribute to other kinematic structures. For every row, an orbit is generated on the Galactic plane; the orbit's kinematics in the SNd is marked cyan in the middle; and the x-y surface of section of the orbit is presented on the right. From top to bottom, we present four orbits corresponding to the kinematics of the Hat, Hyades, and Hercules.}
    \label{fig:futureorbits}
\end{figure*}

Another extension of this study is to extend the influence of orbits in this barred model to other kinematic structures in the SNd. In Fig.\ \ref{fig:futureorbits}, we present a few orbits that may potentially contribute to explaining the other SNd structures. In this figure, each row corresponds to an orbit: the morphology of the orbit is shown in the left; the SNd $L_Z$-$V_R$ kinematics is presented in the middle; and their corresponding surface of section is plotted on the right to identify the orbit families they belong to. From the top, the first row shows a QPO of the outer circular orbit discussed in \S\ref{sec:y0ej}. At this large radius, the non-axisymmetric influence of the bar is very small, so the orbit lives in a nearly axisymmetric environment. This orbit includes the SNd kinematics in the Arch/Hat. These orbits can bring disc stars from a larger radius to the SNd. These outer disc stars are expected to be more Fe-deficient. In \citetalias{LYS2024arXiv241119085L:I}, the Arch/Hat group was found with consistent chemical information to these orbits. 

In the second row, we present a circular QPO that includes the SNd kinematics of the peak of the Hyades group. The surface of section plot identifies it as a QPO of the slow Trojan family. Further below, another peculiar circular QPO is found without a parent periodic orbit. This orbit also includes the kinematics of the Hyades group in the SNd, but unlike the previous orbit, it has a much larger dispersion in $V_R$. Considering the structures in the $L_Z$-$V_R$ kinematic space, the two families of orbits can both contribute to the Hyades group in the SNd.

The bottom row shows a fast Trojan orbit responsible for the Hercules group we discussed extensively in this paper. A few QPOs of the -4:1 resonant orbit also include $L_Z$ similar to Hyades but with a much larger dispersion in $V_R$. This can potentially be related to the origin of the under-density valley between the high $L_Z$ groups and Hercules. Orbits in the barred dynamics model may potentially be a uniform mechanism that explains most of the major kinematic structures in the SNd. If this is correct, it could be possible to find a distribution in the $(y_0,\ E_J)$ initial condition space whose orbits recreate the distribution of the $L_Z$-$V_R$ kinematics space in the SNd.

We notice the discrepancies between the straight lines mapped from single orbits into the $L_Z$-$V_R$ plane and the observed kinematic group overdensities that preserve arched or curved morphologies concaving to the bottom. There are several potential reasons. The first is the two-dimensional approximation. Since we are considering orbits only in the Galactic plane, vertical behaviours of resonance orbits are neglected, which might influence the distribution of the kinematics. However, \cite{Moreno2021MNRAS.506.4687M} investigated similar orbits in the Galactic plane with different potential models for the halo and the bar. Their mock observations of the resonance orbits form curved arches in the U-V plane instead (e.g. see their Fig. 17). Hence, the 2D approximation would not be the reason for the mismatch. Another way the arches can form is by the superposition of a set of quasi-periodic orbits in the orbit family. Due to the difference in $E_J$ and $y_0$, these orbits share similar but not identical kinematics in the SNd, and can result in the curves and arches in the $L_Z$-$V_R$ plane.

It should be further noted, that our study is based on the Galactic model with a slow, long Galactic bar that has a semi-major axis $a = 5.1$\ kpc and pattern speed $\Omega_b = 40$\ km\ s\inv\ kpc\inv. While this is a Galactic model favoured by most recent studies \citep[e.g.][]{Leung2023MNRAS.519..948L, Drimmel2023A&A...670A..10D, Lucchini2024MNRAS.531L..14L}, different model parameters of the Galactic bar can influence the results. Specifically, while the change of the bar length is less influential directly to the orbit families, a longer bar favours a slower pattern speed which is crucial. The pattern speed $\Omega_b$ of the bar is directly related to the corotation radius $R_{\text{CR}} = \Omega_{\text{RC}}^{-1}(\Omega_b)$, which, due to the decreasing Galactic rotation curve $\Omega_{\text{RC}}(R)$ \citep[see][]{Eilers2019ApJ...871..120E, Zhou2023ApJ...946...73Z:RC}, comes inwards with the increase of the bar pattern speed. This also brings Trojan orbits at corotation and orbits at OLR inward. Therefore, with different pattern speeds, the orbits that pass through the SNd are affected by different resonances. For example, prior to \cite{Wegg2015MNRAS.450.4050W}, most studies assumed a fast short bar that has a smaller corotation radius and a smaller OLR radius which allows OLR-related orbits to reach the SNd and prevents Trojan orbits entering the SNd.

However, regardless of the pattern speed and the resulting corotation radius, the stability of Trojan orbits indicates that they should exist in the Galactic disc in any physical barred model. Moreover, the unique features of these orbits ensure their visibility in the kinematics observations. Therefore, if Trojans visit the SNd but do not account for the Hercules group, they are most likely to be responsible for some other anomaly structures in the SNd. If Trojans cannot reach the SNd, we should be able to find signals of their existence in the kinematic space of other neighbourhoods in the disc. \cite{Bernet2022A&A...667A.116B} and \cite{Lucchini2024MNRAS.531L..14L} explored the kinematic space of other radial and azimuthal neighbourhoods, showing evidence supporting the Trojan nature of Hercules.

Furthermore, most fast bar pattern speeds are derived by either the Tremaine-Weinberg method, comparing gas flow to hydrodynamical models, or inferring from kinematic groups like Hercules or Hyades \citep[see][for a review]{Gerhard2011MSAIS..18..185G}. To consider the origin of Hercules, pattern speeds obtained from kinematic groups are biased towards the chosen model and could at the best instance lead to the conclusion of self-consistency. Those obtained from hydrodynamical simulations are highly dependent on the characteristic features considered. The result could vary from 30 to 60\ km\ s\inv\ kpc\inv in different studies \citep[e.g][]{Englmaier1999MNRAS.304..512E:omega60, Weiner1999ApJ...524..112W:omega42, RodriguezF2008A&A...489..115R:omega30}. The range of results includes both the fast and the slow bar scenarios in the Hercules studies and would not be useful to settle the debate directly. Moreover, the direct method of Tremaine-Weinberg \citep{Tremaine1984ApJ...282L...5T:TW} assumes the continuity equation, and studies have challenged its reliability when applied to gas \citep[e.g][]{Borodina2023MNRAS.524.3437B:TW_method}. Therefore, although the bar pattern speed is an important parameter to which the origins of Hercules and other kinematic groups are sensitive, we have to rely on the current favoured value of $\Omega_b \approx 40$\ km\ s\inv\ kpc\inv after including the long bar \citep{Portail2017MNRAS.465.1621P, Bovy2019MNRAS.490.4740B:omega, Li2022ApJ...925...71L}.

In the debate of the fast versus slow bar model, while the driving mechanisms of the Hercules group are different and well-studied in both models, it should be noted that only one mechanism is supported in one scenario. In the fast bar model, as the OLR radius moves inward and OLR orbits pass through the SNd \citep{Fragkoudi2019MNRAS.488.3324F}, the corotation radius is also reduced and the Trojan orbits cannot move out to reach the SNd. On the other hand, in the slow bar scenario, while the corotation Trojan orbits can pass through the SNd, the OLR radius is far beyond the SNd and orbits in the SNd are unlikely to be influenced by the OLR.

In addition, in both models, random radial migration due to perturbations from transient spiral arms can transport inner Galactic orbits out to the SNd and show the kinematics of Hercules stars. However, the stochasticity of this mechanism suggests that these stars are more likely to show a more uniform or Gaussian distribution in the $L_Z$-$V_R$ plane. The super-metal rich stars found in the high $L_Z$ kinematic groups in \citetalias{LYS2024arXiv241119085L:I} can be migrated stars from the inner Galaxy. However, distinguishing the contribution of the Hercules group from Trojans or radial migration is challenging with our numerical solvers. Instead, self-consistent N-body simulations with automatic orbit classifiers would be a more suitable method to consider this issue in future studies.

\section{Conclusions}
\label{sec:summary}

In this paper, we investigated the Trojan scenario of the underlying mechanism of the Hercules group with a barred galaxy model with a long slow bar that reflects our current best understanding of the Galactic bar. By combining our analysis of stellar chemical data from GALAH DR4 and APOGEE DR17 in \citetalias{LYS2024arXiv241119085L:I}, we arrive at the following conclusions:

\begin{itemize}
    \item Two periodic orbit families are associated with the L4 Lagrange point, at corotation resonance with the Galactic bar. One of the families is a near-circular fast family that cuts through contour lines of the effective potential, while the other is a slow family that follows the banana-shaped contour line around the L4. While both families are not allowed to visit the SNd directly, their stability allows them to form very extended quasi-periodic orbits that can pass through the SNd. In the SNd, these QPOs include the $L_Z$-$V_R$ kinematics of the Hercules group with a bias towards positive $V_R$. 

    \item The effective potential $\Phi_{\text{eff}}$ in the corotating frame of the Galactic bar shows the topology of a volcano, the rim of which, consists of the four non-central Lagrange points at corotation radius, prevents the communication between the inner and outer regions. Trojan orbits always cross the rim and provide communication between the inner and outer Galaxy. Hence, Hercules stars with an inner Galactic chemical origin, as indicated in \citetalias{LYS2024arXiv241119085L:I}, are more likely to be captured by the Trojan orbits. As they can bring inner Galactic stars across the rim to the outside and pass through the SNd.

    \item By applying tools of surface of section, the Trojan orbits are found to be highly stable and cover a significant volume in the phase space. Through mock observations, we find that QPOs with higher $E_J$ hold more extended QPOs which include lower $L_Z$ in the SNd. Thus, high $L_Z$ Hercules subgroups are more likely to be associated with low $E_J$ Trojan QPOs and low $L_Z$ Hercules structures were likely related to high $E_J$ Trojan QPOs.

    \item We reveal that the orbits in our barred model can contribute to explaining other kinematic overdensity structures in SNd. Early results associated circular orbits with the LSR kinematics; QPOs of circular orbits from the outer disc with the Arch/hat group; and some peculiar circular QPOs related to the Trojan orbit families with the Hyades group.
\end{itemize}

In conclusion, chemical analysis in \citetalias{LYS2024arXiv241119085L:I} and orbital analysis in this paper indicate that stars in the Hercules group are born in the inner Galaxy and transported over the rim of the effective potential to the SNd. In a slow, long bar scenario, the Trojan orbits at corotation resonance provide a stable ordered mechanism that is capable of carrying inner Galactic stars over the rim and passing through the SNd with the observed Hercules kinematics, especially the lower $L_Z$ and the biased $V_R$.

By considering a more realistic 3D potential, our future work will explore possible origins of the super thin bar, investigate possible scenarios of the capture of inner Galactic stars into Trojan orbits, and explore the 3D behaviours of orbits in the barred Galaxy. We also plan to make use of the stability of Trojan orbits among most physical scenarios to provide further limits to the pattern speed of the Galactic bar.


\section*{Acknowledgements}
We acknowledge the traditional owners of the land on which the AAT and ANU stand, the Gamilaraay, the Ngunnawal and Ngambri people. We pay our respects to elders past, present, and emerging and are proud to continue their tradition of surveying the night sky in the Southern hemisphere.

HJ thanks the Queensland University School of Mathematics and Physics for hospitality and funding support as part of their distinguished visitor program. LYS is grateful to friendly discussions with O. Gerhard during his visit to the Australian National University. Some codes are adopted from course materials by C. Federrath and S. Buder.

We thank the editors and the anonymous referee for the constructive feedback. We acknowledge OUP and MNRAS for kindly granting the APC waiver.

\section*{Software}
The research for this publication was coded in \texttt{Python} v. 3.12 and included packages \texttt{Astropy} v. 6.0.0 \citep{astropy:2013,astropy:2018,astropy:2022}, \texttt{IPython} v. 6.27.1 \citep{ipython:PER-GRA:2007}, \texttt{Matplotlib} v. 3.8.2 \citep{Hunter:2007:matplotlib}, \texttt{NumPy} v. 1.26.4 \citep{harris2020:numpy}, \texttt{SciPy} v. 1.11.4 \citep{2020:SciPy}, and \texttt{tqdm} v. 4.66.1 \citep{tqdm:2022zndo....595120D}.

\section*{Data Availability}

The Python code with the Ferrers barred potential is available in \url{https://github.com/HYuLeeP/FerrersModel}. Code used for data analysis and plotting are available on request. \textit{Gaia} DR3 \citep{Gaia2023A&A...674A...1G} can be obtained via \href{https://gea.esac.esa.int/archive/}{\textit{Gaia} arcive}.
 

\bibliographystyle{mnras}
\bibliography{bibfile} 

\begin{thebibliography}{}
\makeatletter
\relax
\def\mn@urlcharsother{\let\do\@makeother \do\$\do\&\do\#\do\^\do\_\do\%\do\~}
\def\mn@doi{\begingroup\mn@urlcharsother \@ifnextchar [ {\mn@doi@} {\mn@doi@[]}}
\def\mn@doi@[#1]#2{\def\@tempa{#1}\ifx\@tempa\@empty \href {http://dx.doi.org/#2} {doi:#2}\else \href {http://dx.doi.org/#2} {#1}\fi \endgroup}
\def\mn@eprint#1#2{\mn@eprint@#1:#2::\@nil}
\def\mn@eprint@arXiv#1{\href {http://arxiv.org/abs/#1} {{\tt arXiv:#1}}}
\def\mn@eprint@dblp#1{\href {http://dblp.uni-trier.de/rec/bibtex/#1.xml} {dblp:#1}}
\def\mn@eprint@#1:#2:#3:#4\@nil{\def\@tempa {#1}\def\@tempb {#2}\def\@tempc {#3}\ifx \@tempc \@empty \let \@tempc \@tempb \let \@tempb \@tempa \fi \ifx \@tempb \@empty \def\@tempb {arXiv}\fi \@ifundefined {mn@eprint@\@tempb}{\@tempb:\@tempc}{\expandafter \expandafter \csname mn@eprint@\@tempb\endcsname \expandafter{\@tempc}}}

\bibitem[\protect\citeauthoryear{{Abdurro'uf} et~al.,}{{Abdurro'uf} et~al.}{2022}]{APOGEE2022ApJS..259...35A}
{Abdurro'uf} et~al., 2022, \mn@doi [\apjs] {10.3847/1538-4365/ac4414}, \href {https://ui.adsabs.harvard.edu/abs/2022ApJS..259...35A} {259, 35}

\bibitem[\protect\citeauthoryear{{Antoja} et~al.,}{{Antoja} et~al.}{2014}]{Antoja2014A&A...563A..60A}
{Antoja} T.,  et~al., 2014, \mn@doi [\aap] {10.1051/0004-6361/201322623}, \href {https://ui.adsabs.harvard.edu/abs/2014A&A...563A..60A} {563, A60}

\bibitem[\protect\citeauthoryear{{Asano}, {Fujii}, {Baba}, {B{\'e}dorf}, {Sellentin}  \& {Portegies Zwart}}{{Asano} et~al.}{2020}]{Asano2020MNRAS.499.2416A}
{Asano} T.,  {Fujii} M.~S.,  {Baba} J.,  {B{\'e}dorf} J.,  {Sellentin} E.,   {Portegies Zwart} S.,  2020, \mn@doi [\mnras] {10.1093/mnras/staa2849}, \href {https://ui.adsabs.harvard.edu/abs/2020MNRAS.499.2416A} {499, 2416}

\bibitem[\protect\citeauthoryear{{Asplund}, {Grevesse}, {Sauval}  \& {Scott}}{{Asplund} et~al.}{2009}]{Asplund2009ARA&A..47..481A}
{Asplund} M.,  {Grevesse} N.,  {Sauval} A.~J.,   {Scott} P.,  2009, \mn@doi [\araa] {10.1146/annurev.astro.46.060407.145222}, \href {https://ui.adsabs.harvard.edu/abs/2009ARA&A..47..481A} {47, 481}

\bibitem[\protect\citeauthoryear{{Astropy Collaboration} et~al.,}{{Astropy Collaboration} et~al.}{2013}]{astropy:2013}
{Astropy Collaboration} et~al., 2013, \mn@doi [\aap] {10.1051/0004-6361/201322068}, \href {https://ui.adsabs.harvard.edu/abs/2013A&A...558A..33A} {558, A33}

\bibitem[\protect\citeauthoryear{{Astropy Collaboration} et~al.,}{{Astropy Collaboration} et~al.}{2018}]{astropy:2018}
{Astropy Collaboration} et~al., 2018, \mn@doi [\aj] {10.3847/1538-3881/aabc4f}, \href {https://ui.adsabs.harvard.edu/abs/2018AJ....156..123A} {156, 123}

\bibitem[\protect\citeauthoryear{{Astropy Collaboration} et~al.,}{{Astropy Collaboration} et~al.}{2022}]{astropy:2022}
{Astropy Collaboration} et~al., 2022, \mn@doi [\apj] {10.3847/1538-4357/ac7c74}, \href {https://ui.adsabs.harvard.edu/abs/2022ApJ...935..167A} {935, 167}

\bibitem[\protect\citeauthoryear{{Athanassoula}}{{Athanassoula}}{1992}]{Athanassoula1992MNRAS.259..328A}
{Athanassoula} E.,  1992, \mn@doi [\mnras] {10.1093/mnras/259.2.328}, \href {https://ui.adsabs.harvard.edu/abs/1992MNRAS.259..328A} {259, 328}

\bibitem[\protect\citeauthoryear{{Athanassoula}, {Bienayme}, {Martinet}  \& {Pfenniger}}{{Athanassoula} et~al.}{1983}]{Athanassoula1983A&A...127..349A}
{Athanassoula} E.,  {Bienayme} O.,  {Martinet} L.,   {Pfenniger} D.,  1983, \aap, \href {https://ui.adsabs.harvard.edu/abs/1983A&A...127..349A} {127, 349}

\bibitem[\protect\citeauthoryear{{Athanassoula}, {Romero-G{\'o}mez}, {Bosma}  \& {Masdemont}}{{Athanassoula} et~al.}{2009}]{Athanassoula2009MNRAS.400.1706A}
{Athanassoula} E.,  {Romero-G{\'o}mez} M.,  {Bosma} A.,   {Masdemont} J.~J.,  2009, \mn@doi [\mnras] {10.1111/j.1365-2966.2009.15583.x}, \href {https://ui.adsabs.harvard.edu/abs/2009MNRAS.400.1706A} {400, 1706}

\bibitem[\protect\citeauthoryear{{Barros}, {P{\'e}rez-Villegas}, {L{\'e}pine}, {Michtchenko}  \& {Vieira}}{{Barros} et~al.}{2020}]{Barros2020ApJ...888...75B}
{Barros} D.~A.,  {P{\'e}rez-Villegas} A.,  {L{\'e}pine} J. R.~D.,  {Michtchenko} T.~A.,   {Vieira} R. S.~S.,  2020, \mn@doi [\apj] {10.3847/1538-4357/ab59d1}, \href {https://ui.adsabs.harvard.edu/abs/2020ApJ...888...75B} {888, 75}

\bibitem[\protect\citeauthoryear{{Bensby}, {Oey}, {Feltzing}  \& {Gustafsson}}{{Bensby} et~al.}{2007}]{Bensby_Feltzing2007ApJ...655L..89B}
{Bensby} T.,  {Oey} M.~S.,  {Feltzing} S.,   {Gustafsson} B.,  2007, \mn@doi [\apjl] {10.1086/512014}, \href {https://ui.adsabs.harvard.edu/abs/2007ApJ...655L..89B} {655, L89}

\bibitem[\protect\citeauthoryear{{Bernet}, {Ramos}, {Antoja}, {Famaey}, {Monari}, {Al Kazwini}  \& {Romero-G{\'o}mez}}{{Bernet} et~al.}{2022}]{Bernet2022A&A...667A.116B}
{Bernet} M.,  {Ramos} P.,  {Antoja} T.,  {Famaey} B.,  {Monari} G.,  {Al Kazwini} H.,   {Romero-G{\'o}mez} M.,  2022, \mn@doi [\aap] {10.1051/0004-6361/202244070}, \href {https://ui.adsabs.harvard.edu/abs/2022A&A...667A.116B} {667, A116}

\bibitem[\protect\citeauthoryear{{Binney}}{{Binney}}{2020}]{Binney2020MNRAS.495..895B}
{Binney} J.,  2020, \mn@doi [\mnras] {10.1093/mnras/staa1103}, \href {https://ui.adsabs.harvard.edu/abs/2020MNRAS.495..895B} {495, 895}

\bibitem[\protect\citeauthoryear{{Binney} \& {Spergel}}{{Binney} \& {Spergel}}{1982}]{Binney1982ApJ...252..308B}
{Binney} J.,  {Spergel} D.,  1982, \mn@doi [\apj] {10.1086/159559}, \href {https://ui.adsabs.harvard.edu/abs/1982ApJ...252..308B} {252, 308}

\bibitem[\protect\citeauthoryear{{Binney} \& {Tremaine}}{{Binney} \& {Tremaine}}{2008}]{BinneyTremaine2008gady.book.....B}
{Binney} J.,  {Tremaine} S.,  2008, {Galactic Dynamics: Second Edition}

\bibitem[\protect\citeauthoryear{{Binney}, {Gerhard}  \& {Hut}}{{Binney} et~al.}{1985}]{Binney1985MNRAS.215...59B}
{Binney} J.,  {Gerhard} O.~E.,   {Hut} P.,  1985, \mn@doi [\mnras] {10.1093/mnras/215.1.59}, \href {https://ui.adsabs.harvard.edu/abs/1985MNRAS.215...59B} {215, 59}

\bibitem[\protect\citeauthoryear{{Borodina}, {Williams}, {Sormani}, {Meidt}  \& {Schinnerer}}{{Borodina} et~al.}{2023}]{Borodina2023MNRAS.524.3437B:TW_method}
{Borodina} O.,  {Williams} T.~G.,  {Sormani} M.~C.,  {Meidt} S.,   {Schinnerer} E.,  2023, \mn@doi [\mnras] {10.1093/mnras/stad2068}, \href {https://ui.adsabs.harvard.edu/abs/2023MNRAS.524.3437B} {524, 3437}

\bibitem[\protect\citeauthoryear{{Bovy}, {Leung}, {Hunt}, {Mackereth}, {Garc{\'\i}a-Hern{\'a}ndez}  \& {Roman-Lopes}}{{Bovy} et~al.}{2019}]{Bovy2019MNRAS.490.4740B:omega}
{Bovy} J.,  {Leung} H.~W.,  {Hunt} J. A.~S.,  {Mackereth} J.~T.,  {Garc{\'\i}a-Hern{\'a}ndez} D.~A.,   {Roman-Lopes} A.,  2019, \mn@doi [\mnras] {10.1093/mnras/stz2891}, \href {https://ui.adsabs.harvard.edu/abs/2019MNRAS.490.4740B} {490, 4740}

\bibitem[\protect\citeauthoryear{Brent}{Brent}{1973}]{Brent1973}
Brent R.~P.,  1973, {Algorithms for Minimization without Derivatives}, 1st edn.
Prentice-Hall, Englewood Cliffs, New Jersey

\bibitem[\protect\citeauthoryear{{Buder} et~al.,}{{Buder} et~al.}{2024}]{Buder2024arXiv240919858B}
{Buder} S.,  et~al., 2024, \mn@doi [arXiv e-prints] {10.48550/arXiv.2409.19858}, \href {https://ui.adsabs.harvard.edu/abs/2024arXiv240919858B} {p. arXiv:2409.19858}

\bibitem[\protect\citeauthoryear{{Chiba} \& {Sch{\"o}nrich}}{{Chiba} \& {Sch{\"o}nrich}}{2021}]{Chiba2021MNRAS.505.2412C_tree_ring}
{Chiba} R.,  {Sch{\"o}nrich} R.,  2021, \mn@doi [\mnras] {10.1093/mnras/stab1094}, \href {https://ui.adsabs.harvard.edu/abs/2021MNRAS.505.2412C} {505, 2412}

\bibitem[\protect\citeauthoryear{{Chiba}, {Friske}  \& {Sch{\"o}nrich}}{{Chiba} et~al.}{2021}]{Chiba2021MNRAS.500.4710C_resonace}
{Chiba} R.,  {Friske} J. K.~S.,   {Sch{\"o}nrich} R.,  2021, \mn@doi [\mnras] {10.1093/mnras/staa3585}, \href {https://ui.adsabs.harvard.edu/abs/2021MNRAS.500.4710C} {500, 4710}

\bibitem[\protect\citeauthoryear{{Cincotta} \& {Sim{\'o}}}{{Cincotta} \& {Sim{\'o}}}{2000}]{Cincotta2000A&AS..147..205C}
{Cincotta} P.~M.,  {Sim{\'o}} C.,  2000, \mn@doi [\aaps] {10.1051/aas:2000108}, \href {https://ui.adsabs.harvard.edu/abs/2000A&AS..147..205C} {147, 205}

\bibitem[\protect\citeauthoryear{{D'Onghia} \& {L. Aguerri}}{{D'Onghia} \& {L. Aguerri}}{2020}]{DOnghia2020ApJ...890..117D}
{D'Onghia} E.,  {L. Aguerri} J.~A.,  2020, \mn@doi [\apj] {10.3847/1538-4357/ab6bd6}, \href {https://ui.adsabs.harvard.edu/abs/2020ApJ...890..117D} {890, 117}

\bibitem[\protect\citeauthoryear{{Dehnen}}{{Dehnen}}{2000}]{Dehnen2000AJ....119..800D}
{Dehnen} W.,  2000, \mn@doi [\aj] {10.1086/301226}, \href {https://ui.adsabs.harvard.edu/abs/2000AJ....119..800D} {119, 800}

\bibitem[\protect\citeauthoryear{{Drimmel} et~al.,}{{Drimmel} et~al.}{2023}]{Drimmel2023A&A...670A..10D}
{Drimmel} R.,  et~al., 2023, \mn@doi [\aap] {10.1051/0004-6361/202244605}, \href {https://ui.adsabs.harvard.edu/abs/2023A&A...670A..10D} {670, A10}

\bibitem[\protect\citeauthoryear{{Eggen}}{{Eggen}}{1983}]{Eggen1983AJ.....88..642E}
{Eggen} O.~J.,  1983, \mn@doi [\aj] {10.1086/113352}, \href {https://ui.adsabs.harvard.edu/abs/1983AJ.....88..642E} {88, 642}

\bibitem[\protect\citeauthoryear{{Eggen}}{{Eggen}}{1996}]{Eggen1996AJ....112.1595E}
{Eggen} O.~J.,  1996, \mn@doi [\aj] {10.1086/118126}, \href {https://ui.adsabs.harvard.edu/abs/1996AJ....112.1595E} {112, 1595}

\bibitem[\protect\citeauthoryear{{Eilers}, {Hogg}, {Rix}  \& {Ness}}{{Eilers} et~al.}{2019}]{Eilers2019ApJ...871..120E}
{Eilers} A.-C.,  {Hogg} D.~W.,  {Rix} H.-W.,   {Ness} M.~K.,  2019, \mn@doi [\apj] {10.3847/1538-4357/aaf648}, \href {https://ui.adsabs.harvard.edu/abs/2019ApJ...871..120E} {871, 120}

\bibitem[\protect\citeauthoryear{{Englmaier} \& {Gerhard}}{{Englmaier} \& {Gerhard}}{1999}]{Englmaier1999MNRAS.304..512E:omega60}
{Englmaier} P.,  {Gerhard} O.,  1999, \mn@doi [\mnras] {10.1046/j.1365-8711.1999.02280.x}, \href {https://ui.adsabs.harvard.edu/abs/1999MNRAS.304..512E} {304, 512}

\bibitem[\protect\citeauthoryear{{Famaey}, {Jorissen}, {Luri}, {Mayor}, {Udry}, {Dejonghe}  \& {Turon}}{{Famaey} et~al.}{2005}]{Famaey2005A&A...430..165F:stream}
{Famaey} B.,  {Jorissen} A.,  {Luri} X.,  {Mayor} M.,  {Udry} S.,  {Dejonghe} H.,   {Turon} C.,  2005, \mn@doi [\aap] {10.1051/0004-6361:20041272}, \href {https://ui.adsabs.harvard.edu/abs/2005A&A...430..165F} {430, 165}

\bibitem[\protect\citeauthoryear{{Ferrers}}{{Ferrers}}{1877}]{Ferrers1877QJPAM..14....1F}
{Ferrers} N.~M.,  1877, The Quarterly Journal of Pure and Applied Mathematics, \href {https://ui.adsabs.harvard.edu/abs/1877QJPAM..14....1F} {14, 1}

\bibitem[\protect\citeauthoryear{{Fragkoudi} et~al.,}{{Fragkoudi} et~al.}{2019}]{Fragkoudi2019MNRAS.488.3324F}
{Fragkoudi} F.,  et~al., 2019, \mn@doi [\mnras] {10.1093/mnras/stz1875}, \href {https://ui.adsabs.harvard.edu/abs/2019MNRAS.488.3324F} {488, 3324}

\bibitem[\protect\citeauthoryear{{Freeman}}{{Freeman}}{1966}]{Freeman1966MNRAS.133...47F}
{Freeman} K.~C.,  1966, \mn@doi [\mnras] {10.1093/mnras/133.1.47}, \href {https://ui.adsabs.harvard.edu/abs/1966MNRAS.133...47F} {133, 47}

\bibitem[\protect\citeauthoryear{{GRAVITY Collaboration} et~al.,}{{GRAVITY Collaboration} et~al.}{2019}]{GRAVITY2019A&A...625L..10G}
{GRAVITY Collaboration} et~al., 2019, \mn@doi [\aap] {10.1051/0004-6361/201935656}, \href {https://ui.adsabs.harvard.edu/abs/2019A&A...625L..10G} {625, L10}

\bibitem[\protect\citeauthoryear{{GRAVITY Collaboration} et~al.,}{{GRAVITY Collaboration} et~al.}{2021}]{GRAVITY2021A&A...647A..59G}
{GRAVITY Collaboration} et~al., 2021, \mn@doi [\aap] {10.1051/0004-6361/202040208}, \href {https://ui.adsabs.harvard.edu/abs/2021A&A...647A..59G} {647, A59}

\bibitem[\protect\citeauthoryear{{Gaia Collaboration} et~al.,}{{Gaia Collaboration} et~al.}{2023}]{Gaia2023A&A...674A...1G}
{Gaia Collaboration} et~al., 2023, \mn@doi [\aap] {10.1051/0004-6361/202243940}, \href {https://ui.adsabs.harvard.edu/abs/2023A&A...674A...1G} {674, A1}

\bibitem[\protect\citeauthoryear{{Gerhard}}{{Gerhard}}{2011}]{Gerhard2011MSAIS..18..185G}
{Gerhard} O.,  2011, \mn@doi [Memorie della Societa Astronomica Italiana Supplementi] {10.48550/arXiv.1003.2489}, \href {https://ui.adsabs.harvard.edu/abs/2011MSAIS..18..185G} {18, 185}

\bibitem[\protect\citeauthoryear{Harris et~al.,}{Harris et~al.}{2020}]{harris2020:numpy}
Harris C.~R.,  et~al., 2020, \mn@doi [Nature] {10.1038/s41586-020-2649-2}, 585, 357

\bibitem[\protect\citeauthoryear{{Henon} \& {Heiles}}{{Henon} \& {Heiles}}{1964}]{Henon1964AJ.....69...73H}
{Henon} M.,  {Heiles} C.,  1964, \mn@doi [\aj] {10.1086/109234}, \href {https://ui.adsabs.harvard.edu/abs/1964AJ.....69...73H} {69, 73}

\bibitem[\protect\citeauthoryear{{Hunt} \& {Bovy}}{{Hunt} \& {Bovy}}{2018}]{Hunt2018MNRAS.477.3945H_OLR}
{Hunt} J. A.~S.,  {Bovy} J.,  2018, \mn@doi [\mnras] {10.1093/mnras/sty921}, \href {https://ui.adsabs.harvard.edu/abs/2018MNRAS.477.3945H} {477, 3945}

\bibitem[\protect\citeauthoryear{{Hunt}, {Hong}, {Bovy}, {Kawata}  \& {Grand}}{{Hunt} et~al.}{2018}]{Hunt2018MNRAS.481.3794H_Transient}
{Hunt} J. A.~S.,  {Hong} J.,  {Bovy} J.,  {Kawata} D.,   {Grand} R. J.~J.,  2018, \mn@doi [\mnras] {10.1093/mnras/sty2532}, \href {https://ui.adsabs.harvard.edu/abs/2018MNRAS.481.3794H} {481, 3794}

\bibitem[\protect\citeauthoryear{Hunter}{Hunter}{2007}]{Hunter:2007:matplotlib}
Hunter J.~D.,  2007, \mn@doi [Computing in Science \& Engineering] {10.1109/MCSE.2007.55}, 9, 90

\bibitem[\protect\citeauthoryear{{Hunter} et~al.,}{{Hunter} et~al.}{2024}]{Hunter2024arXiv240318000H}
{Hunter} G.~H.,  et~al., 2024, \mn@doi [arXiv e-prints] {10.48550/arXiv.2403.18000}, \href {https://ui.adsabs.harvard.edu/abs/2024arXiv240318000H} {p. arXiv:2403.18000}

\bibitem[\protect\citeauthoryear{{Kalnajs}}{{Kalnajs}}{1991}]{Kalnajs1991dodg.conf..323K}
{Kalnajs} A.~J.,  1991, in {Sundelius} B.,  ed., Dynamics of Disc Galaxies. p.~323

\bibitem[\protect\citeauthoryear{{Khoperskov} \& {Gerhard}}{{Khoperskov} \& {Gerhard}}{2022}]{Khoperskov2022A&A...663A..38K}
{Khoperskov} S.,  {Gerhard} O.,  2022, \mn@doi [\aap] {10.1051/0004-6361/202141836}, \href {https://ui.adsabs.harvard.edu/abs/2022A&A...663A..38K} {663, A38}

\bibitem[\protect\citeauthoryear{{Kim}, {Seo}  \& {Kim}}{{Kim} et~al.}{2012}]{Kim2012ApJ...758...14K_usingferrers}
{Kim} W.-T.,  {Seo} W.-Y.,   {Kim} Y.,  2012, \mn@doi [\apj] {10.1088/0004-637X/758/1/14}, \href {https://ui.adsabs.harvard.edu/abs/2012ApJ...758...14K} {758, 14}

\bibitem[\protect\citeauthoryear{{Leung}, {Bovy}, {Mackereth}, {Hunt}, {Lane}  \& {Wilson}}{{Leung} et~al.}{2023}]{Leung2023MNRAS.519..948L}
{Leung} H.~W.,  {Bovy} J.,  {Mackereth} J.~T.,  {Hunt} J. A.~S.,  {Lane} R.~R.,   {Wilson} J.~C.,  2023, \mn@doi [\mnras] {10.1093/mnras/stac3529}, \href {https://ui.adsabs.harvard.edu/abs/2023MNRAS.519..948L} {519, 948}

\bibitem[\protect\citeauthoryear{{Li}, {Shen}, {Gerhard}  \& {Clarke}}{{Li} et~al.}{2022}]{Li2022ApJ...925...71L}
{Li} Z.,  {Shen} J.,  {Gerhard} O.,   {Clarke} J.~P.,  2022, \mn@doi [\apj] {10.3847/1538-4357/ac3823}, \href {https://ui.adsabs.harvard.edu/abs/2022ApJ...925...71L} {925, 71}

\bibitem[\protect\citeauthoryear{{Li}, {Freeman}, {Jerjen}, {Buder}, {Hayden}  \& {Mondal}}{{Li} et~al.}{2024}]{LYS2024arXiv241119085L:I}
{Li} Y.,  {Freeman} K.,  {Jerjen} H.,  {Buder} S.,  {Hayden} M.,   {Mondal} A.,  2024, \mn@doi [arXiv e-prints] {10.48550/arXiv.2411.19085}, \href {https://ui.adsabs.harvard.edu/abs/2024arXiv241119085L} {p. arXiv:2411.19085}

\bibitem[\protect\citeauthoryear{{Liang}, {Yoon}, {Zhao}, {Li}, {Zhang}  \& {Wu}}{{Liang} et~al.}{2023}]{Liang2023ApJ...956..146L}
{Liang} X.,  {Yoon} S.-J.,  {Zhao} J.,  {Li} Z.,  {Zhang} J.,   {Wu} Y.,  2023, \mn@doi [\apj] {10.3847/1538-4357/acf295}, \href {https://ui.adsabs.harvard.edu/abs/2023ApJ...956..146L} {956, 146}

\bibitem[\protect\citeauthoryear{{Lucchini}, {D'Onghia}  \& {Aguerri}}{{Lucchini} et~al.}{2024}]{Lucchini2024MNRAS.531L..14L}
{Lucchini} S.,  {D'Onghia} E.,   {Aguerri} J. A.~L.,  2024, \mn@doi [\mnras] {10.1093/mnrasl/slae024}, \href {https://ui.adsabs.harvard.edu/abs/2024MNRAS.531L..14L} {531, L14}

\bibitem[\protect\citeauthoryear{{Michtchenko}, {L{\'e}pine}, {P{\'e}rez-Villegas}, {Vieira}  \& {Barros}}{{Michtchenko} et~al.}{2018}]{Michtchenko2018ApJ...863L..37M}
{Michtchenko} T.~A.,  {L{\'e}pine} J. R.~D.,  {P{\'e}rez-Villegas} A.,  {Vieira} R. S.~S.,   {Barros} D.~A.,  2018, \mn@doi [\apjl] {10.3847/2041-8213/aad804}, \href {https://ui.adsabs.harvard.edu/abs/2018ApJ...863L..37M} {863, L37}

\bibitem[\protect\citeauthoryear{{Monari}, {Kawata}, {Hunt}  \& {Famaey}}{{Monari} et~al.}{2017}]{Monari2017MNRAS.466L.113M}
{Monari} G.,  {Kawata} D.,  {Hunt} J. A.~S.,   {Famaey} B.,  2017, \mn@doi [\mnras] {10.1093/mnrasl/slw238}, \href {https://ui.adsabs.harvard.edu/abs/2017MNRAS.466L.113M} {466, L113}

\bibitem[\protect\citeauthoryear{{Monari}, {Famaey}, {Siebert}, {Wegg}  \& {Gerhard}}{{Monari} et~al.}{2019}]{Monari2019A&A...626A..41M}
{Monari} G.,  {Famaey} B.,  {Siebert} A.,  {Wegg} C.,   {Gerhard} O.,  2019, \mn@doi [\aap] {10.1051/0004-6361/201834820}, \href {https://ui.adsabs.harvard.edu/abs/2019A&A...626A..41M} {626, A41}

\bibitem[\protect\citeauthoryear{{Moreno}, {Fern{\'a}ndez-Trincado}, {Schuster}, {P{\'e}rez-Villegas}  \& {Chaves-Velasquez}}{{Moreno} et~al.}{2021}]{Moreno2021MNRAS.506.4687M}
{Moreno} E.,  {Fern{\'a}ndez-Trincado} J.~G.,  {Schuster} W.~J.,  {P{\'e}rez-Villegas} A.,   {Chaves-Velasquez} L.,  2021, \mn@doi [\mnras] {10.1093/mnras/stab1908}, \href {https://ui.adsabs.harvard.edu/abs/2021MNRAS.506.4687M} {506, 4687}

\bibitem[\protect\citeauthoryear{{Murray} \& {Dermott}}{{Murray} \& {Dermott}}{1999}]{Murray1999ssd..book.....M}
{Murray} C.~D.,  {Dermott} S.~F.,  1999, {Solar System Dynamics}, \mn@doi{10.1017/CBO9781139174817.
}

\bibitem[\protect\citeauthoryear{{Pagel} \& {Edmunds}}{{Pagel} \& {Edmunds}}{1981}]{Pagel1981ARA&A..19...77P}
{Pagel} B.~E.~J.,  {Edmunds} M.~G.,  1981, \mn@doi [\araa] {10.1146/annurev.aa.19.090181.000453}, \href {https://ui.adsabs.harvard.edu/abs/1981ARA&A..19...77P} {19, 77}

\bibitem[\protect\citeauthoryear{{Perek}}{{Perek}}{1962}]{Perek1962AdA&A...1..165P}
{Perek} L.,  1962, \mn@doi [Advances in Astronomy and Astrophysics] {10.1016/B978-1-4831-9919-1.50008-X}, \href {https://ui.adsabs.harvard.edu/abs/1962AdA&A...1..165P} {1, 165}

\bibitem[\protect\citeauthoryear{P\'erez \& Granger}{P\'erez \& Granger}{2007}]{ipython:PER-GRA:2007}
P\'erez F.,  Granger B.~E.,  2007, \mn@doi [Computing in Science and Engineering] {10.1109/MCSE.2007.53}, 9, 21

\bibitem[\protect\citeauthoryear{{P{\'e}rez-Villegas}, {Portail}, {Wegg}  \& {Gerhard}}{{P{\'e}rez-Villegas} et~al.}{2017}]{PerezVillegas2017ApJ...840L...2P}
{P{\'e}rez-Villegas} A.,  {Portail} M.,  {Wegg} C.,   {Gerhard} O.,  2017, \mn@doi [\apjl] {10.3847/2041-8213/aa6c26}, \href {https://ui.adsabs.harvard.edu/abs/2017ApJ...840L...2P} {840, L2}

\bibitem[\protect\citeauthoryear{{Portail}, {Gerhard}, {Wegg}  \& {Ness}}{{Portail} et~al.}{2017}]{Portail2017MNRAS.465.1621P}
{Portail} M.,  {Gerhard} O.,  {Wegg} C.,   {Ness} M.,  2017, \mn@doi [\mnras] {10.1093/mnras/stw2819}, \href {https://ui.adsabs.harvard.edu/abs/2017MNRAS.465.1621P} {465, 1621}

\bibitem[\protect\citeauthoryear{{Quillen} et~al.,}{{Quillen} et~al.}{2018}]{Quillen2018MNRAS.478..228Q}
{Quillen} A.~C.,  et~al., 2018, \mn@doi [\mnras] {10.1093/mnras/sty865}, \href {https://ui.adsabs.harvard.edu/abs/2018MNRAS.478..228Q} {478, 228}

\bibitem[\protect\citeauthoryear{{Rodriguez-Fernandez} \& {Combes}}{{Rodriguez-Fernandez} \& {Combes}}{2008}]{RodriguezF2008A&A...489..115R:omega30}
{Rodriguez-Fernandez} N.~J.,  {Combes} F.,  2008, \mn@doi [\aap] {10.1051/0004-6361:200809644}, \href {https://ui.adsabs.harvard.edu/abs/2008A&A...489..115R} {489, 115}

\bibitem[\protect\citeauthoryear{{Sellwood} \& {Wilkinson}}{{Sellwood} \& {Wilkinson}}{1993}]{Sellwood1993RPPh...56..173S}
{Sellwood} J.~A.,  {Wilkinson} A.,  1993, \mn@doi [Reports on Progress in Physics] {10.1088/0034-4885/56/2/001}, \href {https://ui.adsabs.harvard.edu/abs/1993RPPh...56..173S} {56, 173}

\bibitem[\protect\citeauthoryear{{Sormani}, {Binney}  \& {Magorrian}}{{Sormani} et~al.}{2015}]{Sormani2015MNRAS.454.1818S}
{Sormani} M.~C.,  {Binney} J.,   {Magorrian} J.,  2015, \mn@doi [\mnras] {10.1093/mnras/stv2067}, \href {https://ui.adsabs.harvard.edu/abs/2015MNRAS.454.1818S} {454, 1818}

\bibitem[\protect\citeauthoryear{{Sormani} et~al.,}{{Sormani} et~al.}{2022}]{Sormani2022MNRAS.512.1857S}
{Sormani} M.~C.,  et~al., 2022, \mn@doi [\mnras] {10.1093/mnras/stac639}, \href {https://ui.adsabs.harvard.edu/abs/2022MNRAS.512.1857S} {512, 1857}

\bibitem[\protect\citeauthoryear{{Tremaine} \& {Weinberg}}{{Tremaine} \& {Weinberg}}{1984}]{Tremaine1984ApJ...282L...5T:TW}
{Tremaine} S.,  {Weinberg} M.~D.,  1984, \mn@doi [\apjl] {10.1086/184292}, \href {https://ui.adsabs.harvard.edu/abs/1984ApJ...282L...5T} {282, L5}

\bibitem[\protect\citeauthoryear{Virtanen et~al.,}{Virtanen et~al.}{2020}]{2020:SciPy}
Virtanen P.,  et~al., 2020, \mn@doi [Nature Methods] {10.1038/s41592-019-0686-2}, \href {https://rdcu.be/b08Wh} {17, 261}

\bibitem[\protect\citeauthoryear{{Wegg}, {Gerhard}  \& {Portail}}{{Wegg} et~al.}{2015}]{Wegg2015MNRAS.450.4050W}
{Wegg} C.,  {Gerhard} O.,   {Portail} M.,  2015, \mn@doi [\mnras] {10.1093/mnras/stv745}, \href {https://ui.adsabs.harvard.edu/abs/2015MNRAS.450.4050W} {450, 4050}

\bibitem[\protect\citeauthoryear{{Weiner} \& {Sellwood}}{{Weiner} \& {Sellwood}}{1999}]{Weiner1999ApJ...524..112W:omega42}
{Weiner} B.~J.,  {Sellwood} J.~A.,  1999, \mn@doi [\apj] {10.1086/307786}, \href {https://ui.adsabs.harvard.edu/abs/1999ApJ...524..112W} {524, 112}

\bibitem[\protect\citeauthoryear{{Wheeler}, {Abril-Cabezas}, {Trick}, {Fragkoudi}  \& {Ness}}{{Wheeler} et~al.}{2022}]{Wheeler2022ApJ...935...28W}
{Wheeler} A.,  {Abril-Cabezas} I.,  {Trick} W.~H.,  {Fragkoudi} F.,   {Ness} M.,  2022, \mn@doi [\apj] {10.3847/1538-4357/ac7da0}, \href {https://ui.adsabs.harvard.edu/abs/2022ApJ...935...28W} {935, 28}

\bibitem[\protect\citeauthoryear{{Zhou}, {Li}, {Huang}  \& {Zhang}}{{Zhou} et~al.}{2023}]{Zhou2023ApJ...946...73Z:RC}
{Zhou} Y.,  {Li} X.,  {Huang} Y.,   {Zhang} H.,  2023, \mn@doi [\apj] {10.3847/1538-4357/acadd9}, \href {https://ui.adsabs.harvard.edu/abs/2023ApJ...946...73Z} {946, 73}

\bibitem[\protect\citeauthoryear{{da Costa-Luis} et~al.,}{{da Costa-Luis} et~al.}{2024}]{tqdm:2022zndo....595120D}
{da Costa-Luis} C.,  et~al., 2024, {tqdm: A fast, Extensible Progress Bar for Python and CLI}, \mn@doi{10.5281/zenodo.595120}

\bibitem[\protect\citeauthoryear{{de Vaucouleurs} \& {Freeman}}{{de Vaucouleurs} \& {Freeman}}{1972}]{Vaucouleurs1972VA.....14..163D}
{de Vaucouleurs} G.,  {Freeman} K.~C.,  1972, \mn@doi [Vistas in Astronomy] {10.1016/0083-6656(72)90026-8}, \href {https://ui.adsabs.harvard.edu/abs/1972VA.....14..163D} {14, 163}

\makeatother
\end{thebibliography}

\appendix
\section{The Ferrers bar potential}
\label{app:fbar}
This section presents the method used to calculated the ferrers spheroidal bar potential model.

\subsection{The potential}
The density of the Ferrer's spheroidal bar is

\begin{equation*}
    \rho = \rho_0 (1-m^2)^2, \quad \text{where}\quad m(x,y) = \frac{x^2}{a^2}+\frac{y^2}{c^2},
\end{equation*}

for $m\leq 1$, where $a>c$ and $\epsilon^2 = a^2-c^2$. Define $\psi(x,y)$ by,

\begin{align}
    \label{eqn:mg1}
    y^2 \tan^2\psi +x^2\sin^2\psi&=\epsilon^2 &(m>1),\\
    \cos\psi &= \frac{c}{a} &(m<1).
    \label{eqn:ml1}
\end{align}

Define

\begin{equation}
    W_{lk}=2\int_0^\psi \tan^{2l-1}\theta\ \sin^{2k-1}\theta\ d\theta.
    \label{eqn:wlk}
\end{equation}

Write $x=\Tilde{x}\epsilon$, $y=\Tilde{y}\epsilon$, $a=\Tilde{a}\epsilon$, then the potential is
\begin{equation*}
    \Phi = \frac{\pi G \rho_0 ac^2}{\epsilon}\Tilde{\Phi},
\end{equation*}
where, by dropping tildes,
\begin{multline*}
    \Tilde{\Phi} = \frac{1}{3}W_{10}-(y^2W_{20}+x^2W_{11})+(y^4W_{30}+2y^2x^2W_{21}+x^4W_{12})\\
    -\frac{1}{3}(y^6W_{40}+3y^4x^2W_{31}+3y^2x^4W_{22}+x^6W_{13}).
\end{multline*}

The mass of the bar is $M = 32\pi\rho_0 ac^2/105$. Let the bar rotate with angular velocity $\Omega$. Write

\begin{equation*}
    Q = \frac{105GM}{32\Omega^2\epsilon^3}.
\end{equation*}

Then
\begin{equation*}
    \Phi=Q\Omega^2\epsilon^2\Tilde{\Phi}.
\end{equation*}

\newcommand{\parfrac}[2]{\frac{\partial #1}{\partial #2}}
\newcommand{\ww}[1]{W_{#1}}

\subsection{The potential gradient}
Within the bar ($m < 1$), $W+{lk}$s are constant and the potential gradient is
\begin{align*}
    \frac{\partial\Tilde{\Phi}}{\partial x} = & -2xW_{11}+(4xy^2W_{21}+4x^3W_{12})\\&\qquad-(2xy^4W_{31}+4x^3y^2W_{22}+2x^5W_{13}),\\
    \frac{\partial\Tilde{\Phi}}{\partial y} = & -2yW_{20}+(4y^3W_{30}+4x^2yW_{21})\\&\qquad-(2y^5W_{40}+4x^2y^3W_{31}+2x^4yW_{22}).
\end{align*}

Outside the bar ($m>1$), $W_{lk}=W_{lk}(x,y)$ are functions of $x \text{ and } y$. Then,

\begin{align*}
    \frac{\partial\Tilde{\Phi}}{\partial x} = & \frac{1}{3}\frac{\partial W_{10}}{\partial x}-(y^2\frac{\partial W_{20}}{\partial x}+xW_{11}+x^2\frac{\partial W_{11}}{\partial x})\\
    &+(y^4\frac{\partial W_{30}}{\partial x}+4xy^2W_{21}+2x^2y^2\frac{\partial W_{21}}{\partial x}+x^4\frac{\partial W_{12}}{\partial x}+4x^3 W_{12})\\
    &-\frac{1}{3}(y^6\parfrac{\ww{40}}{x}+6xy^4\ww{31}+3x^2y^4\parfrac{\ww{31}}{x}+3x^4y^2\parfrac{\ww{22}}{x}\\
    &\hspace{2cm}+12x^3y^2\ww{22}+x^6\parfrac{\ww{13}}{x}+6x^5\ww{13})
\end{align*}

\begin{align*}
    \parfrac{\Tilde{\Phi}}{y} = & \frac{1}{3}\parfrac{\ww{10}}{y}-(2y\ww{20}+y^2\parfrac{\ww{20}}{y}+x^2\parfrac{\ww{11}}{y})\\
    &+(4y^3\ww{30}+y^4\parfrac{\ww{30}}{y}+4x^2y\ww{21}+2x^2y^2\parfrac{\ww{21}}{y}+x^4\parfrac{\ww{12}}{y})\\
    &-\frac{1}{3}(6y^5\ww{40}+y^6\parfrac{\ww{40}}{y}+12x^2y^3\ww{31}+3x^2y^4\parfrac{\ww{31}}{y}\\
    &\hspace{2cm}+6x^4y\ww{22}+3x^4y^2\parfrac{\ww{22}}{y}+x^6\parfrac{\ww{13}}{y})
\end{align*}

\subsection{Computing the potential gradient}

In equation \ref{eqn:wlk}, substitute $u = \sin \theta$; $v(x, y) = \sin \psi$. Then,
\begin{equation}
    \label{eqn:wlk_uv}
    \ww{lk}=2\int_0^v\frac{u^{2l+2k-2}\ du}{(1-u^2)^l}.
\end{equation}

From equations \ref{eqn:mg1} and \ref{eqn:ml1},

\begin{align}
    \label{eqn:ellpse}
    x^2v^2(1-v^2)+y^2v^2+v^2 &=1& \text{for } m > 1,\\
    v &= \frac{1}{a}& \text{for } m \leq 1.
\end{align}

Equation \ref{eqn:ellpse} is an ellipse in ($x,y$) with semi-axes ($1/v$) and $\sqrt{(1+v^2)}/v$. For $v = a$, the ellipse is the edge of the Ferrer's bar.

The derivatives of $\ww{lk}$ are,
\begin{align*}
    \parfrac{\ww{lk}}{x} &= 2 \frac{v^{2l+2k-2}}{(1-v^2)^l}\parfrac{v}{x}, \qquad \text{where,}\\
    \parfrac{v}{x}&=\frac{-xv(1-v^2)}{x^2(1-2v^2)+y^2+1},\\
    \parfrac{v}{y}&=\frac{-yv}{x^2(1-2v^2)+y^2+1}.
\end{align*}

The package \texttt{scipy.optimise.brentq} \citep{Brent1973} is adopted to derive v from equation \ref{eqn:ellpse}.

\subsection{Recurrence relations for W}

Start from equation \ref{eqn:wlk_uv}. Define 6 integrals:
\begin{equation*}
    J_n = 2\int \frac{v^{2n}dv}{(1-v^2)^{n+2}} \qquad I_n = 2\int v^{2n}dv,
\end{equation*}
for $n = 0,\ 1,\ 2$,

\begin{align*}
    J_0 &= \frac{1}{2}(\frac{2v}{1-v^2}+\log\frac{1+v}{1-v}),\\
    J_1 &= \frac{1}{8}(\frac{2v(1+v^2)}{(1-v^2)^2}-\log\frac{1+v}{1-v}),\\
    J_2 &= \frac{1}{48}(\frac{2v(3v^4+8v^2-3)}{(1-v^2)3}+\log\frac{1+v}{1-v}).
\end{align*}

We obtain $\ww{10}$ and $\ww{40}$ directly,
\begin{align*}
    \ww{10} &= \log\frac{1+v}{1-v},\\
    \ww{40} &= \frac{66v^4 -80v^3 -15(1-v^2)^3\ww{10}+30v}{48(1-v^2)^3}.
\end{align*}

Then the remaining terms can be obtained by recurrence relations,
\begin{align*}
    \ww{20} &= J_0 - \ww{10},\\
    \ww{30} &= J_1 - \ww{20},\\
    \ww{40} &= J_2 - \ww{30},\\
    \ww{11} &= \ww{10} - I_0,\\
    \ww{12} &= \ww{11} - I_1,\\
    \ww{13} &= \ww{12} - I_2,\\
    \ww{21} &= \ww{20} - \ww{11},\\
    \ww{22} &= \ww{21} - \ww{12},\\
    \ww{31} &= \ww{30} - \ww{21}.
\end{align*}

\section{The Evolution of Surface of Section space}
\label{app:sos_evo}
In this section, surfaces of section with $E_J$ from $-0.85$ to $0.00$. Throughout all $E_J$, the inner circular retrograde family shown in Figures \ref{fig:sos_high}f and \ref{fig:sos_low}f is the most stable family. At super high $E_J$ this family expands on the surfaces, showing significant increases in stability. The increased energy and stability also result in the appearance of more island orbits in the family. At $E_J$ below $\Phi_{\text{L2}}$, the outer retrograde circular family becomes the dominant family in the outer disc. As seen in Fig.\ \ref{fig:futureorbits}, the outer circular family can be associated with the most regular planar circular motions in the disc. Apart from these, the Trojan families, especially the fast Trojan family, cover a large area on the surface, indicating their identity as the next most stable orbit after the previous circular orbits.

\begin{figure*}
    \centering
    \includegraphics[height=.93\textheight]{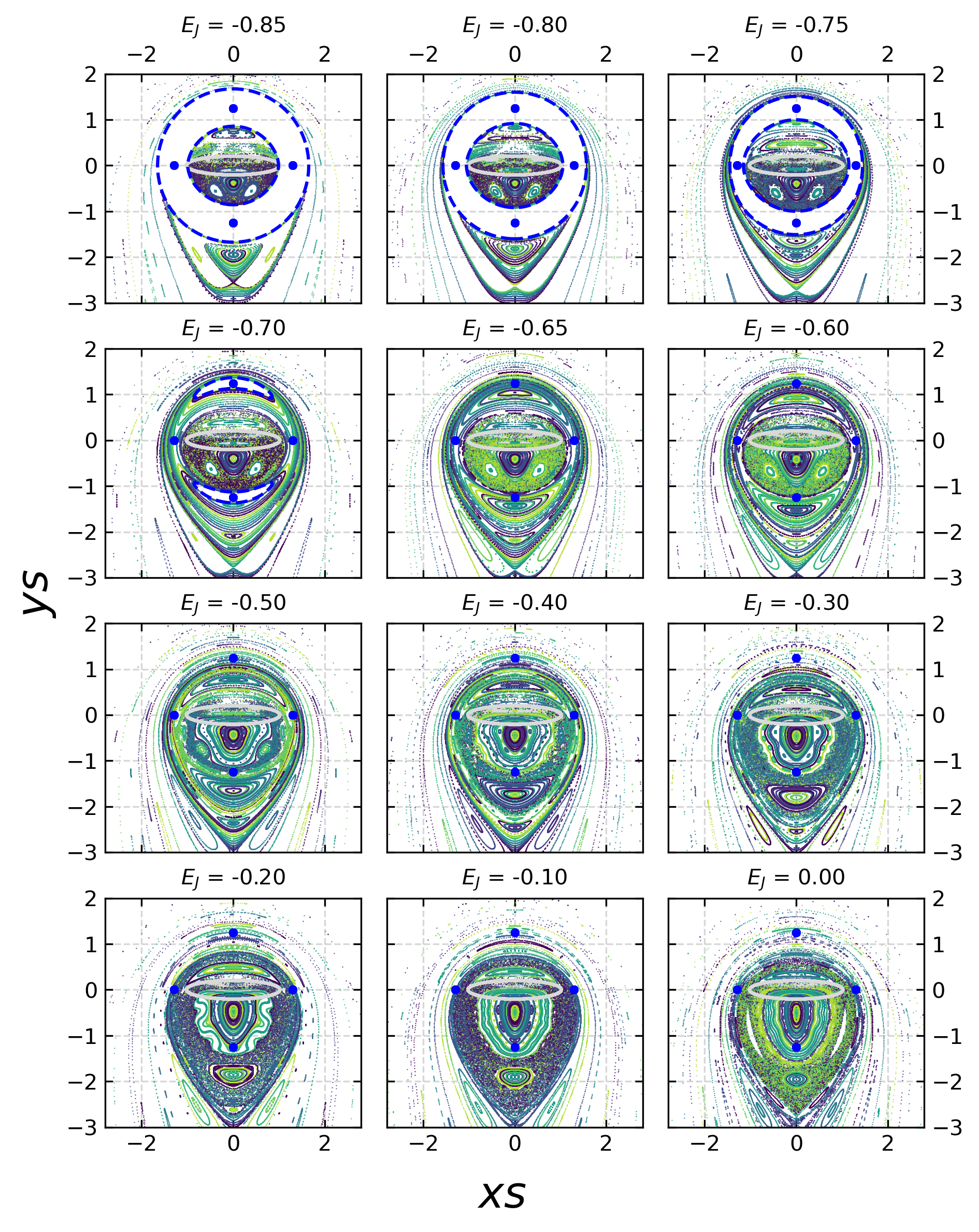}
    \caption{The variation of orbit families on the $x$-$y$ surface of section over $E_J$ from $E_J = -0.85$ to $E_J = 0.00$. The fast Trojan orbit family covers a large extent in the phase space and exists over a large range of $E_J$s. The central retrograde circular family is the most stable family that exist across all the $E_J$ presented.}
    \label{fig:sos_ej}
\end{figure*}

\bsp	
\label{lastpage}
\end{document}